\documentstyle[aps,prl,twocolumn,epsfig]{revtex}
\begin{document}
%%%%%%%%%%%%%%%%%%%%%%%%%%%%%%%%%%%%%%%%%%
\newcommand{\beqa}{\begin{eqnarray}}
\newcommand{\eeqa}{\end{eqnarray}}
\newcommand{\beq}{\begin{equation}}
\newcommand{\eeq}{\end{equation}}
\newcommand{\dg}{\dagger}
\newcommand{\sig}{\sigma}
\newcommand{\vektor}[1]{\mbox{\boldmath $#1$}}
\newcommand{\eff}{\mbox{\scriptsize{eff}}}
\newcommand{\crit}{\mbox{\scriptsize{crit}}}
\newcommand{\doex}{\mbox{\scriptsize{de}}}
%%%%%%%%%%%%%%%%%%%%%%%%%%%%%%%%%%%%%%%%%%
\twocolumn[\hsize\textwidth\columnwidth\hsize\csname
@twocolumnfalse\endcsname
\title{Ordering of localized moments in Kondo lattice models}
\author{Graeme Honner and Mikl\'{o}s Gul\'{a}csi}
\address{Department of Theoretical Physics, 
 Institute of Advanced Studies \\
The Australian National University,
Canberra, ACT 0200, Australia}
\date{\today}
\maketitle
\begin{abstract}
We describe the transition from a ferromagnetic phase, 
to a disordered paramagnetic phase, which occurs in 
one-dimensional 
Kondo lattice models with partial conduction band 
filling. The transition is the quantum 
order-disorder transition of the transverse-field 
Ising chain, and reflects double-exchange ordered 
regions of localized spins being gradually 
destroyed as the coupling to the 
conduction electrons is reduced. For incommensurate 
conduction band filling, the low-energy properties 
of the localized spins near the transition are 
dominated by anomalous ordered (disordered) 
regions of localized spins which survive into the 
paramagnetic (ferromagnetic) phase. 
Many interesting properties follow, including a 
diverging susceptibility for a finite range of  
couplings into the paramagnetic phase. Our critical 
line equation, together with numerically determined 
transition points, are used to determine the effective 
range of the double-exchange interaction. Models 
considered are the spin 1/2 Kondo lattices with 
ferromagnetic and antiferromagnetic couplings,
and the Kondo lattice  
with repulsive interactions between the conduction  
electrons. 
\end{abstract}
]

\section{INTRODUCTION}

The Kondo lattice model (KLM) describes the interaction 
between a conduction band and a lattice of localized 
spins. The hamiltonian for the KLM is 
\beq
H=-t \sum_{<i,j>}\sum_{\sig} c_{i\sig}^{\dg}c_{j\sig}^{} +
J\sum_{j} {\bf S}_{c j} {\bf \cdot} {\bf S}_{j}\, ,
\label{klm}
\eeq
where $t>0$ is the conduction electron 
(or simply electron) hopping, and
$<i,j>$ denotes nearest neighbors. 
${\bf S}_{j}$ are spin $1/2$ operators 
for the localized spins, and 
${\bf S}_{cj}=\frac{1}{2}\sum_{\sig,\sig '}
c_{j\sig}^{\dg} {\vektor{\sig}}_{\sig,\sig '} c_{j\sig '}^{}$ are 
pseudospin operators for the conduction electrons. 
${\vektor{\sig}}$ are Pauli spin matrices, and 
$c^{}_{j\sig},c^{\dg}_{j\sig}$ the electron  
site operators. We choose units so that the hopping $t = 1$, and 
measure the Kondo coupling $J$ in units of $t$.
The KLM is an effective model 
for heavy-fermion systems when the coupling $J$ is 
small and positive \cite{Lee}, and models colossal 
magnetoresistance (CMR) materials when $J$ is large and 
negative \cite{Zang}. 

In the following, we give a comprehensive description 
of the transition from a ferromagnetic (FM) ordering of 
the localized spins at stronger couplings, to a 
quantum disordered paramagnetic (PM) phase at weaker 
couplings. The transition has been identified in the 
one-dimensional (1D) KLM with partial conduction band 
filling both analytically \cite{Honner}, and 
in numerical simulations by a variety of methods 
\cite{Troyer,Tsunetsugu,Moukouri95,Caprara,Yunoki}. 
We focus mainly on the $J > 0$ 1D KLM 
relevant to heavy-fermion systems, in which the 
localized spins model $f$-electrons in lanthanide 
or actinide compounds. The bulk of the numerical simulations 
\cite{Troyer,Tsunetsugu,Moukouri95,Caprara} 
are devoted to this model. 
Partial conduction band filling $n = N_{c}/N < 1$ is 
assumed throughout, where $N_{c}$ is the 
number of electrons, and $N$ the number of 
localized spins. 
Our approach 
generalizes to 1D KLMs with repulsive 
interactions between the electrons, and 
also to the KLM with a FM coupling $J < 0$. These 
models are considered in Sec.\ IV.

A summary of our method is as follows: 
Abelian bosonization is used 
to describe the conduction band. 
Using a unitary transformation, 
the bosonized KLM is written in terms of a basis of 
states in which the localized spin and electron spin  
degrees of freedom are 
coupled. In the new basis, the competing interactions 
leading to the FM-PM transition are clearly exhibited. 
The competing effects are double-exchange FM ordering 
at stronger coupling, and spin-flip disorder processes 
at weaker coupling. Since the FM-PM transition is 
signalled by the ordering of the localized spins,  
we take expectation values for the electron 
Bose fields, and obtain an effective hamiltonian for 
the localized spins. The effective hamiltonian maps 
to a transverse-field Ising chain close to the 
FM-PM phase boundary, and we determine the critical 
line for the resulting quantum order-disorder transition, 
as well as many properties of the localized spins near 
criticality. At weak-coupling deep in the PM phase, 
the effective hamiltonian determines 
RKKY-like behavior, with dominant correlations in the 
localized spins at $2k_{F}$ of the 
conduction band. 

The new element here is an emphasis on the
double-exchange interaction, which tends to 
align the localized spins at stronger couplings. 
Double-exchange is often ignored in discussions of the 
$J > 0$ KLM. Following its introduction 
20 years ago by Doniach \cite{Doniach}, 
the $J > 0$ KLM has usually been discussed in terms of 
a competition between Kondo singlet formation, and the
RKKY interaction. For the KLM with a half-filled conduction 
band $n = 1$, this characterization appears sufficient. 
For the KLM with a partially-filled conduction band $n < 1$, it has   
been realised that neither Kondo singlet formation, nor 
the RKKY interaction, are sufficient to describe rigorously established 
properties of the model \cite{Yanagisawa}. A succession 
of analytic \cite{one-electron,strong-coupling,Honner} 
and numerical \cite{Troyer,Tsunetsugu,Moukouri95,Caprara}  
results have established an 
extensive region of FM ordering at stronger 
coupling. This cannot be explained in terms of 
RKKY, which operates at weak-coupling, nor in 
terms of Kondo singlets, since they are non-magnetic. The missing 
element is double-exchange ordering due to an 
excess of localized spins over conduction electrons. 
Double-exchange requires only that $N > N_{c}$ (i.e.\ 
$n < 1$), and that the conduction electron hopping $t \neq 0$. It 
operates in any dimension, for any sign of 
the coupling $J$, and for any magnitude $S_{j}^{2}$ 
of the localized spins \cite{Anderson}. 
The double-exchange
interaction is specific to the Kondo lattice, and 
is absent in single- or dilute-impurity 
systems in which the situation is reversed, and the 
electrons greatly outnumber the localized spins. 

Double-exchange is conceptually a very simple 
interaction: For $n < 1$ each electron has on average 
more than one localized spin to screen, and 
consequently hops between several adjacent spins 
gaining screening energy at each site, together with 
a gain in kinetic energy. Since hopping is 
energetically most favorable for electrons which preserve 
their spin as they hop (called coherent hopping), 
this tends to align the 
underlying localized spins \cite{Anderson}. 
We show in Sec.\ II that coherent conduction 
electron hopping over a characteristic length $\alpha$ 
may be incorporated into a bosonization description which 
keeps the electrons finitely delocalized. At lengths 
beyond $\alpha$, the electrons 
are described by collective density fluctuations,  
as is usual in 1D Fermi systems. 
The electrons remain finitely delocalized  
over shorter lengths, and describe coherent hopping over several 
adjacent sites. This tends to align  
the underlying localized spins at stronger coupling. 
$\alpha$ measures the effective 
range of the double-exchange interaction, and is 
in principle a function of both filling $n$ and 
coupling $J$.

Our approach generates a ground-state phase diagram for 
the partially-filled 1D KLM in agreement with available 
exact and numerical results for the $J > 0$ KLM 
\cite{one-electron,strong-coupling,Troyer,Tsunetsugu,Moukouri95,Caprara}, 
for the $J > 0$ KLM with repulsive interactions between 
the conduction electrons \cite{Yanagisawa94,Moukouri96}, 
and for the KLM with a FM coupling $J < 0$ 
\cite{Yunoki}. Interesting new properties we determine  
include the following: \\ 
(i) For 
incommensurate conduction band filling, anomalous regions 
of double-exchange ordered localized spins survive close to the 
transition in the PM phase. Similarly, anomalous regions 
of disorder survive close to the transition in the 
FM phase. These regions exist due to the inability of the 
incommensurate conduction band to either totally order, 
or totally disorder the localized spins as the transition is 
crossed. Although the anomalous regions are very dilute, 
they dominate the low-energy properties of the localized spins. 
Many interesting results follow, including a diverging 
susceptibility in the PM phase for a finite range of 
couplings close to the transition. \\
(ii) The effective range of double-exchange ordering,  
measured by the length $\alpha$ for coherent  
electron hopping, is determined  
using our critical line equation together with numerically 
determined transition points. We find that $\alpha$ 
scales as $1/\sqrt{|J|}$ for any sign of the coupling $J$ 
on the transition line. For $J < 0$, 
$\alpha \rightarrow 0$ as half-filling is approached;  
double-exchange is ineffective as $n \rightarrow 1$, and the
transition line diverges. For $J > 0$, $\alpha$ is reduced to 
less than a lattice spacing as $n \rightarrow 1$, but remains 
non-zero; the transition line remains finite close  
to half-filling for the $J > 0$ KLM.

An outline of the paper is as follows: In Sec.\ II we 
derive the effective hamiltonian for the localized spins 
in the partially-filled $J > 0$ 1D KLM, 
and provide a justification of results 
briefly reported elsewhere \cite{Honner}. In Sec.\ III 
the effective hamiltonian is analyzed to determine the 
ground-state magnetic phase diagram, together with
low-energy properties of the localized spins near the resulting 
FM-PM transition. Using our results, 
together with available numerical data, 
we determine the effective range of the double-exchange 
interaction on the transition line. Our method 
may be generalized to describe the 
FM-PM transition in other KLMs, and in
Sec.\ IV we consider the 1D KLM with 
interactions between the  
electrons, and show that the FM-PM transition 
is pushed to lower values of $J$ for repulsive 
interactions. We consider also the KLM with a 
FM $J < 0$ coupling,
and follow a similar analysis to that for the 
$J > 0$ case. The ground-state phase 
diagram is determined, together with the effective 
range of the double-exchange 
interaction on the resulting FM-PM transition line. 
The paper concludes in Sec.\ V with a summary of results, 
and a discussion of the general features of the 
FM-PM transition in partially-filled KLMs.

\section{EFFECTIVE HAMILTONIAN FOR THE LOCALIZED SPINS}

A large class of 1D many-electron systems may 
be described using bosonization techniques 
\cite{Voit}: The electron fields 
may be represented in terms of collective 
density operators which satisfy  
bosonic commutation relations. Bose  
representations provides a non-perturbative
description which, in general, is far easier to manipulate than
a formulation in terms of fermionic operators. In
the 1D KLM, the conduction band may be  
bosonized, but not the localized spins. This is because the 
spins are strictly localized, and their Fermi velocity 
vanishes. Moreover, since there is no direct interaction between 
the localized spins in the KLM, it not possible to use 
bosonization via a direct Jordan-Wigner transformation. 
We therefore bosonize only the conduction band.

Bose representations are conveniently written in terms 
of Bose fields, defined as follows: 
\beqa
\phi_{\nu}(j) &=& -i\sum_{k \neq 0}\frac{\pi}{kL}[\nu_{+}(k) 
 + \nu_{-}(k)] \Lambda_{\alpha}(k)e^{ikja}, 
\nonumber \\ 
\theta_{\nu}(j) &=& -i\sum_{k \neq 0}\frac{\pi}{kL}[\nu_{+}(k) 
 - \nu_{-}(k)] \Lambda_{\alpha}(k)e^{ikja},
\nonumber \\
\Pi_{\nu}(j) &=& \frac{\pi}{L}\sum_{k \neq 0}[\nu_{+}(k) 
 - \nu_{-}(k)] \Lambda_{\alpha}(k)e^{ikja} 
\nonumber \\
 &=& \partial_{x}\theta_{\nu}(j).
\label{bosefields}
\eeqa
In Eqs.\ (\ref{bosefields}) $\nu = \rho, \sig$ labels charge and spin, 
with charge and spin density operators
\beqa
\rho_{r}(k) = \sum_{\sig}\rho_{r\sig}(k),
\quad \, \, 
\sig_{r}(k) = \sum_{\sig} \sig \rho_{r\sig}(k).
\label{rhosig}
\eeqa
$\sig = \pm 1$ as the spin is up or down, respectively.
The density operators $\rho_{r\sig}(k)$ are the basic bosonic objects,  
and are defined by 
\beqa
\rho_{r\sig}(k) = \sum_{0 < rk' < \pi/a} 
c^{\dg}_{k' - \frac{k}{2} \sig} 
c^{}_{k' + \frac{k}{2} \sig}\, .
\label{RHOrsig}
\eeqa 
The density operators describe collective coherent 
particle-hole excitations about 
the right ($r = +$) and left ($r = -$) Fermi points 
at $+k_{F}$ and $-k_{F}$, respectively. ($k_{F} = 
\pi n/2a$ with $a$ 
the lattice spacing. The system 
length is $L = Na$.) The density operators  
$\rho_{r\sig}(k)$ are bosonic 
for wave vectors $|k|$ up to $\alpha^{-1}$. For these 
wave vectors we have \cite{Tomonaga}
\beqa
[\rho_{r\sig}(k), \rho_{r'\sig '}(k')] = 
\delta_{r, r'}\,\delta_{k, -k'}\,\delta_{\sig, \sig '}\, 
\frac{rkL}{2\pi}\, . 
\label{bosecom}
\eeqa
In bosonization, $\alpha$ measures the minimum wavelength for  
the densities $\rho_{r\sig}(k)$ which satisfy the bosonic 
commutation relations Eq.\ (\ref{bosecom}). 
Straightforward calculation, as in Ref.\ [17], 
shows that $\alpha$ must satisfy 
$\alpha \gtrsim {\cal O}(k_{F})^{-1}$. This is clear also on  
physical grounds; fermionic density operators are not 
collective, and hence cannot be bosonic, at wavelengths 
of the order of the average interparticle spacing. 
In the Bose fields of Eqs.\ (\ref{bosefields}), 
$\Lambda_{\alpha}(k)$ is 
a cut-off function on bosonic density operators. The cut-off 
function is an even function of $k$, and satisfies 
$\Lambda_{\alpha}(k) \approx 1$ when $|k| < \alpha^{-1}$,
and $\Lambda_{\alpha}(k) \approx 0$ when $|k| > \alpha^{-1}$. 
$\Lambda_{\alpha}(k)$ ensures that only bosonic density 
operators enter the Bose fields. The commutation relations 
between the Bose fields 
are then c-numbers. The physical significance of the Bose fields 
is as potentials: 
$\partial_{x} \phi_{\nu}(j)$ is proportional to the 
$\nu$-density at site $j$, and $\partial_{x} \theta_{\nu}(j)
= \Pi_{\nu}(j)$ is proportional to the average $\nu$-current 
at $j$. ($\partial_{x}\psi_{\nu}(j)$, $\psi = \phi , \theta$, 
is shorthand for $\partial_{x}\psi_{\nu}(x/a)$ 
evaluated at $x = ja$.) 

Bose representations for Fermi operators are derived by requiring 
that they correctly reproduce the commutation relations of the Fermi 
operators with the density operators $\rho_{r\sig}(k)$, and by 
requiring that they correctly reproduce the non-interacting ($J = 0$) 
expectation values. Since the states generated by the density operators 
$\rho_{r\sig}(k)$ span the 1D state space \cite{Schick}, this prescription 
ensures that the Bose representations will reproduce the same matrix 
elements as the original Fermi operators. 
In this way, the hopping term in Eq.\ (\ref{klm}) is given by
\beqa
H_{0} = \frac{v_{F}a}{4\pi}\sum_{\nu,j}\left\{ \Pi_{\nu}^{2}(j) 
+ [\partial_{x}\phi_{\nu}(j)]^{2}\right\}\, , 
\label{Ho}
\eeqa
with a linearized dispersion, and Fermi velocity 
$v_{F} = 2a\sin(\pi n/2)$ in units of $t$.
To bosonize the KLM, we also require representations for the 
on-site Fermi 
bilinears $c^{\dg}_{j\sig}c^{}_{j\sig'} = 
\sum_{r,r'}c^{\dg}_{rj\sig}c^{}_{r'j\sig'}$. The off-diagonal 
bilinears $c^{\dg}_{rj\sig}c^{}_{r'j\sig'}$, in which $r \neq r'$ 
and/or $\sig \neq \sig'$, may be constructed from the 
Bose representation for the single 
Fermi site operators 
\beqa
c_{rj\sig} &=& \sqrt{Aa/2\alpha}\,\, \exp i[rk_{F}ja 
 + \Psi_{r\sig}(j)], 
\nonumber \\
\Psi_{r\sig}(j) &=& \{ \theta_{\rho}(j) + r\phi_{\rho}(j) 
+ \sig[\theta_{\sig}(j) + r\phi_{\sig}(j)]\}/2,
\label{Crjsigbose}
\eeqa
where $A$ is a dimensionless constant 
depending on the cut-off function 
$\Lambda_{\alpha}(k)$. 
The representation Eq.\ (\ref{Crjsigbose}) for $c_{rj\sig}$ 
reproduces the correct commutation relations only with 
long-wavelength density operators $\rho_{r\sig}(k)$. Thus  
the representation Eq.\ (\ref{Crjsigbose}) 
may not correctly reproduce the short-range properties of the 
original conduction electron site operators $c_{j\sig}$.
The Bose representation for the diagonal on-site 
bilinears, i.e.\ the density operators in real space, may be 
obtained directly from their Fourier expansion: 
\beqa
\sum_{r} c^{\dg}_{rj\sig}c^{}_{rj\sig} = 
\frac{a}{2\pi}\partial_{x}
[\phi_{\rho}(j) + \sig \phi_{\sig}(j)] 
\label{diag}
\eeqa
to an additive constant depending on $n$.
As for $H_{0}$, and in constrast to $c_{rj\sig}$, 
this representation is exact. 
Substituting these representations into  
Eq.\ (\ref{klm}) gives the bosonized KLM hamiltonian
\beqa
H &=& \frac{v_{F}a}{4\pi}\sum_{j,\nu}
\left\{ \Pi_{\nu}^{2}(j) + [\partial_{x}\phi_{\nu}(j)]^{2}
\right\}
\nonumber \\
& + & \frac{Ja}{2\pi}\sum_{j}
[\partial_{x}\phi_{\sig}(j)] S_{j}^{z}
\nonumber \\
&+& A\frac{Ja}{2\alpha}\sum_{j}\left\{
\cos [\phi_{\sig}(j)] + \cos[2k_{F}ja + \phi_{\rho}(j)]\right\}
\nonumber \\
&& \quad \quad \quad \quad \times
\left(e^{-i\theta_{\sig}(j)}S_{j}^{+} + {\rm h.c.}\right) 
\nonumber \\
&-& A\frac{Ja}{\alpha} \sum_{j}\sin[\phi_{\sig}(j)]
\sin[2k_{F}ja + \phi_{\rho}(j)]S_{j}^{z}\, .  
\label{bklm}
\eeqa

The bosonized hamiltonian 
generates the same behavior as the original  
hamiltonian provided that the conduction electrons are not 
strongly localized. In particular, the bosonized hamiltonian 
does not directly describe resonant on-site Kondo scattering: At 
strong-coupling the electrons localize, with each 
forming a Kondo singlet with the localized spin at the same site 
\cite{Hirsch,strong-coupling}. The localized singlet 
formation is governed by the 
short-range properties of the spin-flip bilinears 
$c^{\dg}_{j\sig}c^{}_{j-\sig}$. However, the Bose 
representation for these terms, derived from Eq.\  
(\ref{Crjsigbose}), is reliable only at long-wavelengths.
It describes the properties of a spin-flipped  
electron only at large distances from the scattering site. 
This provides a good description at weaker couplings, as  
usual in the bosonization of 1D Fermi systems, but may be  
insufficient when the coupling is strong enough that the 
electron becomes trapped on-site by the localized spin. 

A second point to note about the bosonized hamiltonian concerns 
spin-rotation symmetry. $SU(2)$ symmetry is manifest in the  
original hamiltonian, Eq.\ (\ref{klm}), for both the 
electrons and the localized spins, but is obscured
in the bosonized version. This is due to 
the use of abelian bosonization, which treats the 
electron spin $z$ direction on a special footing, 
and breaks the $SU(2)$ electron spin-rotation 
symmetry down to $U(1)$. To see the effect of this,
note that the original hamiltonian preserves both 
the total spin $S_{\mbox{\scriptsize{tot}}}$ as well as its 
$z$ component $S^{z}_{\mbox{\scriptsize{tot}}}$, and at
stronger couplings in the FM phase may be
decoupled into subspaces with different values of 
$S^{z}_{\mbox{\scriptsize{tot}}}$. (See, for example, 
Ref.\ [12].)
Abelian bosonization effectively singles out the subspace
with maximal $S^{z}_{\mbox{\scriptsize{tot}}}$ in the
FM phase (cf.\ Eq.\ (\ref{tbklm}) below), and may be
physically motivated by crystal-field effects.

\subsection{Unitary transformation}

A simple method for determining the ordering induced on the 
localized spins by the electrons is to choose a 
basis of states in which competing effects become more 
transparent. This is achieved by applying a unitary 
transformation which changes to a basis of states in which 
the conduction electron spin degrees of freedom are coupled 
directly to the localized spins. We choose the transformation
\beqa
\exp({\rm S})\, , \quad {\rm S}=i \frac{Ja}{2\pi v_{F}}
\sum_{j}\theta_{\sig}(j) \,S_{j}^{z}\, .
\nonumber 
\eeqa
A variant of this transformation was first used by Emery
and Kivelson for the single-impurity Kondo problem 
\cite{Emery},
and was later generalized to the 1D KLM \cite{ZKE}.
The usage here is different. Ref.\ [21] aimed to describe the 
conduction electrons, and the transformation 
was used to remove the spin current 
field $\theta_{\sig}(j)$ from the hamiltonian. Here 
we aim 
to describe interactions between the localized spins, in which  
the electrons act as the mediators. 
The form of the transformation is then chosen so as to make 
explicit a FM ordering of the localized spins. This effect
was entirely missed in the previous work \cite{ZKE}. 
The factor 
$Ja/2\pi v_{F}$ is chosen so that terms of the form 
$[\partial_{x}\phi_{\sig}(j)]S_{j}^{z}$ exactly cancel 
in the transformed hamiltonian. This permits the ground-state 
$S_{j}^{z}$ configuration to be chosen independent of the 
on-site electron spin density in the transformed basis. 

Transformed operators 
$\tilde{O} = e^{-{\rm S}}Oe^{{\rm S}}$ may be 
calculated using the standard commutator expansion
\beqa
\tilde{O} = O + [O,{\rm S}]/1! 
+ [[O,{\rm S}],{\rm S}]/2! + \cdots \; . 
\label{comseries} 
\eeqa 
Using $[S_{j}^{x}, S_{j'}^{y}] = i\delta_{j,j'}S_{j}^{z}$ 
etc., we get 
\beqa
\tilde{S}^{z}_{j} &=&  S^{z}_{j} \, 
\nonumber \\
\tilde{S}^{\pm}_{j} &=&  S^{\pm}_{j}
\exp\left\{\mp i \frac{Ja}{2\pi v_{F}} \theta_{\sig}(j)\right\}
\nonumber
\eeqa
so that S rotates the localized spins in the 
$xy$-plane depending 
on the local electron spin current field. S 
does not alter the $S_{j}^{z}$ configuration.
The only Bose field transformed under S is the  
electron spin density field, which takes a form that
depends on the localized spin configuration all along the chain:
\beqa
\tilde{\phi}_{\sig}(j) &=& \phi_{\sig}(j) + K(j) \, , 
\nonumber 
\eeqa
where $K(j)$ is a long-range object which essentially 
counts all the $S^{z}_{j'}$ to the right of site $j$, and 
subtracts from that the $S^{z}_{j'}$ to the left of $j$:
\beqa
K(j) &=& i \frac{Ja}{2\pi v_{F}} \sum_{j'}
[\phi_{\sig}(j),\theta_{\sig}(j')]S_{j'}^{z} \, .
\label{Kj}
\eeqa
Here the Bose field commutator  
$[\phi_{\sig}(j),\theta_{\sig}(j')] \rightarrow 
{\rm sign}(j - j')\, i\pi$ for $(j - j')a \gg \alpha$, 
and this provides the 
main contribution to $K(j)$. This term is discussed further in 
Sec.\ III. Related to the 
transformation of the spin density field is the transformation 
of the actual electron spin density at $j$. This  
depends on the local $S_{j'}^{z}$ configuration. The 
transformed spin density  
$\widetilde{\partial_{x}\phi}_{\sig}(j)$ is given by 
\beqa
\partial_{x}\phi_{\sig}(j) 
-\frac{Ja}{\pi v_{F}}\sum_{j'}\left\{\int_{0}^{\infty}
dk\,\cos[k(j-j')a]\Lambda_{\alpha}^{2}(k)\right\}S_{j'}^{z},
\nonumber
\eeqa
The integral here is the Bose field 
commutator $i[\phi_{\sig}(j), \Pi_{\sig}(j')]/2$, and is 
discussed further below.

After some manipulation, the above results give the transformed 
KLM hamiltonian
\beqa
\tilde{H} & = &  \frac{v_{F}a}{4\pi}\sum_{\nu,j}
\left\{ \Pi_{\nu}^{2}(j)
[\partial_{x}\phi_{\nu}(j)]^{2} \right\}
\nonumber \\
& - & \frac{J^{2}a^{2}}{4\pi^{2}v_{F}}
\sum_{j,j'}\left\{\int_{0}^{\infty}dk\,\cos[k(j-j')a]
\Lambda_{\alpha}^{2}(k)\right\}S_{j}^{z}S_{j'}^{z}
\nonumber \\
& + & A\frac{Ja}{2\alpha}\sum_{j}
\left\{ \cos [K(j)+\phi_{\sig}(j)]
+ \cos [2k_{F}ja+\phi_{\rho}(j)] \right\}
\nonumber \\
&& \quad \quad \quad \times
\left( e^{-i (1+Ja/2\pi v_{F})
\theta_{\sig}(j)}S_{j}^{+} + {\rm{h.c.}}\right)
\nonumber \\
& - & A\frac{Ja}{\alpha}  \sum_{j}
\sin [K(j)+\phi_{\sig}(j)]
\sin [2k_{F}ja + \phi_{\rho}(j)] S_{j}^{z} \nonumber \\
\label{tbklm}
\eeqa
provided that the cut-off function is not too `soft': 
$\Lambda_{\alpha}^{m}(k) \approx \Lambda_{\alpha}(k), m=2,3,4.$
(Discrepancies near $|k| \approx \alpha^{-1}$ introduce 
negligible corrections.) Note that the unitary transformation 
has been carried out exactly and not perturbatively, 
i.e.\ there 
has been no artificial truncation of the commutator series of 
Eq.\ (\ref{comseries}). (The c-number Bose field commutators 
are essential for this.) It follows that the transformed 
hamiltonian of Eq.\ (\ref{tbklm}) is identical to the bosonized 
hamiltonian, and is an exact rewriting of Eq.\ (\ref{bklm}) in 
terms of a new basis of states in which the conduction band 
and localized spins are interwoven.

\subsection{Double-exchange ordering}

The important new term in the transformed hamiltonian 
Eq.\ (\ref{tbklm}) is the 
second:
\beqa
-  \frac{J^{2}a^{2}}{4\pi^{2}v_{F}}
\sum_{j,j'}\left\{\int_{0}^{\infty}dk\,\cos[k(j-j')a]
\Lambda_{\alpha}^{2}(k)\right\}S_{j}^{z}S_{j'}^{z}\, .
\label{FMterm}
\eeqa
It represents a non-perturbative effective interaction 
between the localized spins, and is the only one of this
type to be derived for the KLM; other 
effective interactions, namely the RKKY interaction 
at weak-coupling, and the strong-coupling effective interaction 
of Sigrist {\it et al.} \cite{strong-coupling}, are both 
perturbative. We consider FM in the KLM in some detail in this 
subsection. First we analyze the properties of 
the interaction described by Eq.\ (\ref{FMterm}), and 
describe how it arises from the bosonization of the 
conduction band. Second we present previously known  
properties of the double-exchange interaction.  
The interaction of Eq.\ (\ref{FMterm}) shares these 
properties, and we identify it as the double-exchange 
interaction in the KLM. Since double-exchange is 
usually not considered in discussions of the KLM 
with $J > 0$, we 
conclude the subsection with a simple intuitive picture 
of double-exchange ordering in the KLM at low 
conduction band filling.

Eq.\ (\ref{FMterm}) possesses the following properties: \\
(i) the term originates, via bosonization and 
then the unitary transformation, from the terms $H_{0}$ 
and the forward scattering part of 
$(J/2) \sum_{j} (n_{j\uparrow} - n_{j\downarrow}) 
S^{z}_{j}$ ($n_{j\sig} = c^{\dg}_{j\sig}c^{}_{j\sig}$) in 
the KLM hamiltonian Eq.\ (\ref{klm}). (Note that the Bose 
representations for the electrons in these terms are exact.) \\
(ii) Eq.\ (\ref{FMterm}) is 
independent of the sign of $J$, and takes the same form  
for any magnitude $S_{j}^{2}$ of the localized spins. \\
(iii) Since Eq.\ (\ref{FMterm}) is of order $J^{2}$, 
whereas the remaining terms in the transformed hamiltonian 
Eq.\ (\ref{tbklm}) are of order $J$, the interaction    
Eq.\ (\ref{FMterm}) dominates the ordering of the localized 
spins as $J$ increases. \\ 
(iv) Eq.\ (\ref{FMterm}) is FM for all 
(differentiable) choices of the cut-off function 
$\Lambda_{\alpha}(k)$.  \\
To give examples of the form of the FM interaction 
Eq.\ (\ref{FMterm}) in real space, consider 
Gaussian and exponential cut-off  
functions defined by 
$\Lambda_{\alpha}(k) = \exp(-\alpha^{2}k^{2}/2)$ 
for a Gaussian cut-off, and by 
$\Lambda_{\alpha}(k) = \exp(-\alpha |k|/2)$ for an 
exponential cut-off. 
For these cut-off functions, the integral in 
Eq.\ (\ref{FMterm}) reduces to
\beqa
\int_{0}^{\infty} dk \,\cos(kja) 
\Lambda^{2}_{\alpha}(k)      
= (\sqrt{\pi}/2\alpha) \exp-(ja/2\alpha)^{2}
\eeqa 
for the Gaussian, and 
\beqa 
\int_{0}^{\infty} dk \,\cos(kja)\Lambda^{2}_{\alpha}(k) 
= \alpha/(\alpha^{2} + (ja)^{2})
\eeqa
for the exponential. The integrals are positive
and non-negligible for $ja \lesssim \alpha$. 
The form of the FM interaction for Gaussian and 
exponential cut-off functions is shown in Fig.\ 1. 
It is clear that the length $\alpha$ 
characterizes the effective range of the FM interaction   
of Eq.\ (\ref{FMterm}). 

\begin{figure}[tb]
\centering
\epsfig{file=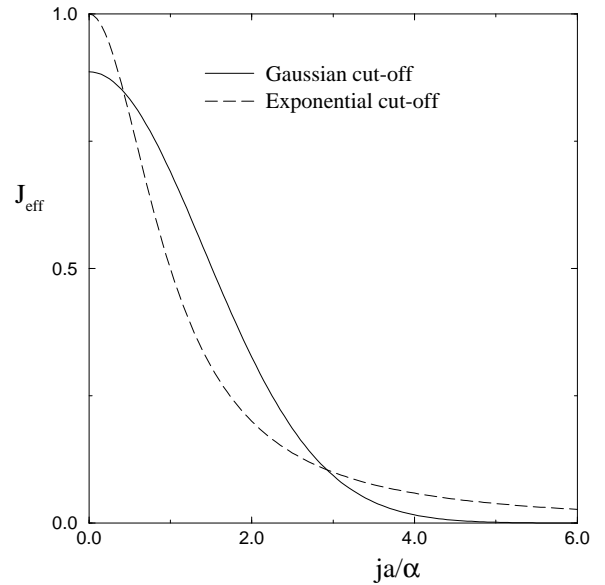,height=9cm}
\caption{The range in real space of the FM interaction 
Eq.\ (\ref{FMterm}) for exponential 
$\exp -(\alpha|k|/2)$ and Gaussian 
$\exp -(\alpha^{2} k^{2}/2)$ cut-off 
functions $\Lambda_{\alpha}(k)$. 
$J_{\eff}$ is the interaction strength 
in units of $\alpha J^{2}a^{2}/4\pi^{2}v_{F}$.}
\end{figure}

The interaction Eq.\ (\ref{FMterm}) 
originates from the bosonization of the  
conduction band as follows: 
At wavelengths beyond $\alpha$, the  
electrons are involved in collective density
fluctuations Eq.\ (\ref{RHOrsig}). These fluctuations 
involve large numbers of electrons, and satisfy bosonic 
commutation relations Eq.\ (\ref{bosecom}). 
This is the standard  
behavior of 1D many-electron systems for weak 
interactions. At wavelengths below 
$\alpha$, the density fluctuations are not  
collective, and do not satisfy bosonic 
commutation relations. Since bosonization describes 
fluctuations only over separations beyond $\alpha$, 
the bosonization description is equivalent to keeping the electrons 
finitely delocalized over $\alpha$, with the electrons 
preserving their spin over this range.
Eq.\ (\ref{FMterm}) is the ordering consequently 
induced on the localized spins by the finitely 
delocalized electrons, and arises formally from  
the Bose field commutator 
$[\phi_{\sig}(j), \Pi_{\sig}(j')]$. This commutator 
takes canonical $\delta$-function form in field theory 
\cite{Voit}, but is 
smeared over a range $\alpha \gtrsim {\cal O}(k_{F})^{-1}$ 
for the conduction band, which has a finite density 
$n$ of electrons. The smearing reflects the 
inability of the Bose fields to distinguish separations 
below $\alpha$.

We turn now to briefly summarize previously known properties of 
the double-exchange interaction. Double-exchange  
was first proposed many years ago by 
Zener \cite{Zener} to explain FM ordering in mixed-valency 
manganites, and is of much current interest in relation to 
CMR materials \cite{Zang}. The essential 
characteristic required for a system to exhibit 
double-exchange ordering is that the number of  
electrons be less than the number of localized 
spins. In this case, consider first infinitely strong 
coupling $J = \infty$. Each electron forms a perfectly 
localized on-site spin singlet (or triplet for $J < 0$) 
with the localized spin at the same site. The remaining 
$N - N_{c}$ unpaired localized spins are free. 
When the conduction electron 
hopping is turned on, the electrons gain energy by hopping 
to unoccupied sites, since they gain energy both by 
screening the unpaired localized spin, together with a 
gain in kinetic energy. Since electrons tend to 
preserve their spin as they hop, called coherent hopping 
\cite{Zang}, this tends to align the 
underlying localized spins \cite{Anderson}. This is the 
double-exchange mechanism, and is generated by the 
conduction electron kinetic energy and 
the diagonal part of the on-site interaction between electrons 
and localized spins. Double-exchange is always FM, and dominates 
at stronger couplings \cite{Zener}. It 
is the physical basis of 
the FM rigorously established in the 1D KLM by 
Sigrist {\it et al.} \cite{one-electron,strong-coupling}. 
(See also Ref.\ [10].) Since double-exchange ordering 
requires only $N > N_{c}$ and a non-vanishing 
hopping, its existence (as opposed to other properties 
such as its effective range)
does not depend on the sign of $J$, 
nor on the magnitude of the localized spins. 
This is clear from the early analysis of Anderson and 
Hasegawa \cite{Anderson}; their result does not
depend on the sign of $J$, except for some numerical 
prefactors, and is semiclassical with nearly trivial 
quantum modifications. 

Properties (i)-(iv) above for the interaction of 
Eq.\ (\ref{FMterm}) are identical to those of a 
double-exchange interaction. This  
leads us to identify Eq.\ (\ref{FMterm}) as the 
double-exchange interaction in the partially-filled 
1D KLM. Coherent conduction electron hopping, which generates 
double-exchange ordering, 
is described in the bosonization by electrons 
finitely delocalized over lengths $\alpha$,  
and $\alpha$ measures the effective 
range of the double-exchange interaction, as in Fig.\ 1. 
Note that $\alpha$ enters the bosonization description 
as an undetermined but finite length; from bosonization, we 
know only that $\alpha \gtrsim {\cal O}(k_{F})^{-1}$, and that 
in general $\alpha$ will be a function both of filling 
$n$ and coupling $J$ \cite{Tomonaga}. 
In Secs.\ III and IV (cf.\ Fig.\ 4), we determine $\alpha$ 
in a special case by using our results together with 
available numerical data. 

Since many discussions of the $J > 0$ KLM neglect the
double-exchange interaction, it is useful to present 
the following simple characterization 
valid at low conduction band filling. 
Double-exchange may be described by the KLM hamiltonian 
Eq.\ (\ref{klm}) with spin-flip interactions ignored:
\beqa
H_{\doex} = -t\sum_{j}\left(c^{\dg}_{j\sig}c^{}_{j+1\sig}
 + {\rm h.c.} \right) 
 + J/2\sum_{j} (n_{j\uparrow} - n_{j\downarrow})S^{z}_{j}\, .
\nonumber
\eeqa
The occupation of a site by an 
electron with the same spin as the localized 
spin costs an energy $J/2$. We exclude 
these states as a first approximation valid at stronger 
couplings. 
At small filling, we consider a finitely 
delocalized electron of spin $\sig$ spread over 
sites $j$ for which the localized spins 
$S^{z}_{j}$ have spin $-\sig$ for $J > 0$.  
From $H_{\doex}$, 
the wavefunction $\psi_{\sig}(x)$ for the electron, in the 
continuum limit, satisfies the nonlinear Schr\"{o}dinger equation
\beqa
\partial^{2}_{x}\psi_{\sig}(x) 
+ (Jm/2)|\psi_{\sig}(x)|^{2} \psi_{\sig}(x) 
= 2 m E\psi_{\sig}(x)
\label{nls}
\eeqa
with $m$ the bare electron mass. The electron gains energy 
due to its occupation $|\psi_{\sig}(x)|^{2}$ of a point with 
localized spin $S^{z}_{x}$ of spin $-\sig$, and 
this generates the non-linearity. Finitely 
delocalized solutions of Eq.\ (\ref{nls}) are solitons 
with \cite{Makhankov}
\beqa
\psi_{\sig}(x) = B\, e^{ix}\,
{\rm sech}\left(B\sqrt{Jm/4}\,(x-x_{0})
\right)\, . 
\label{soliton}
\eeqa
B and $x_{0}$ are constants. Our simplified picture 
of the KLM at low conduction 
band filling is then of a gas of solitons. The solitons may  
be pictured as spin polarons \cite{Holstein,one-electron}, 
and describe the dressing of each 
electron by a cloud of localized spins which align opposite 
to the conduction electron spin for $J > 0$. 
The spatial extension of the polarization  
cloud characterizes the range of the indirect FM ordering 
induced on the localized spins by the electron, 
and is equivalent to the effective range $\alpha$ of the 
double-exchange interaction as described previously. From 
Eq.\ (\ref{soliton}), the polarization cloud decays 
exponentially at large distances with a characteristic 
length scale proportional to $1/\sqrt{J}$. 
This gives a low density 
form $\alpha/a \propto 1/\sqrt{J}$. At vanishingly small 
fillings, we may use the exact solution of Sigrist 
{\it et al.} \cite{one-electron} for the KLM with 
one conduction electron. The polarization cloud  
decays exponentially for small $J$ in the FM phase, with 
a characteristic length $\alpha/a = \sqrt{2/J}$.

Since the interaction Eq.\ (\ref{FMterm}) 
is short-range for all finite $\alpha$ 
(correct for all finite $n$, cf.\ Fig.\ 4), 
we approximate it 
in the usual way by its nearest-neighbor form
$-{\cal J}\sum_{j} S^{z}_{j}S^{z}_{j+1}$, where 
\beqa
{\cal J} = \frac{J^{2}a^{2}}{2\pi^{2}v_{F}} 
\int_{0}^{\infty}dk\, \cos(ka) \Lambda^{2}_{\alpha}(k)\, .
\label{calJ}
\eeqa
We do not expect critical properties to be affected by 
this approximation.

\subsection{Effective hamiltonian}

We aim to use the transformed hamiltonian Eq.\ (\ref{tbklm}) 
to determine the ground-state properties of the  
localized spins in the partially-filled 1D KLM. 
Our concentration on the localized spins is motivated 
by the results of numerical simulations on the KLM. 
Simulations on large chains have been carried out 
on the partially-filled 1D KLM using
quantum Monte Carlo \cite{Troyer}, and using the 
density-matrix renormalization-group (DMRG)  
\cite{Moukouri95,Caprara}. The results uniformly show that the
correlations between the localized spins are
much stronger than the correlations between the conduction 
electrons, and the FM-PM transition is signalled 
by the crossover from FM to incommensurate 
(generally $2k_{F}$) correlations in the structure factor of the 
localized spins. The corresponding electron correlations 
are observed to weakly track those of the 
localized spins, and the electron 
momentum distribution shows no dramatic change as the 
FM-PM transition line is crossed \cite{Moukouri95}. The freezing 
of the electron spin degrees of freedom  
occurs only at very strong coupling deep in the FM phase. 

An effective hamiltonian for the localized spins is  
obtained from Eq.\ (\ref{tbklm}) by taking appropriately 
chosen expectation values for the conduction electron Bose 
fields.  
Since the Bose fields enter only in the weak-coupling 
terms of order $J$ in Eq.\ (\ref{tbklm}), we 
approximate the Bose fields by their non-interacting $J=0$ 
expectation values:
\beqa
\langle \phi_{\nu}(j) \rangle_{0} = 
\langle \theta_{\sig}(j) \rangle_{0} = 0\, .   
\label{approx}
\eeqa
This holds for the charge 
density field $\phi_{\rho}(j)$, since at weak-coupling 
the charge structure factor is free 
electron-like \cite{Troyer}. For the spin fields, Eq.\ 
(\ref{approx}) follows from 
real-space renormalization-group studies \cite{Jullien}, which 
show that the spin degrees of freedom of the 1D KLM flow to the 
non-interacting fixed point at weak-coupling. 
(This property is specific to the Kondo lattice \cite{Jullien}. 
For the single-impurity Kondo model at zero temperature, 
the spin coupling renormalizes to infinity for all 
$J \neq 0$.) 
Note that Eq.\ (\ref{approx}) is further
supported by a study of the 
1D KLM with $t$-$J$ interacting conduction electrons 
\cite{Moukouri96}. Using a combination of exact diagonalization 
and the DMRG, the same ordering was observed to be induced 
on the localized spins as in the pure KLM of Eq.\ (\ref{klm}), 
and confirms the insensitivity of the ordering 
to the details of the conduction electron behavior. 
The transformed hamiltonian Eq.\ (\ref{tbklm}) now 
reduces to an effective hamiltonian for the localized spins: 
\beqa
H_{\eff} &=& -{\cal J}\sum_{j}S^{z}_{j}S^{z}_{j+1} 
\nonumber \\
&+& A\frac{Ja}{\alpha}\sum_{j} 
\{ \cos[K(j)] + \cos[2k_{F}ja]\}S^{x}_{j}
\nonumber \\
&-& A\frac{Ja}{\alpha}\sum_{j}
\sin[K(j)] \sin[2k_{F}ja]S^{z}_{j}\, . 
\label{Heff}
\eeqa

\section{GROUND-STATE MAGNETIC PHASE DIAGRAM}

In this section the effective hamiltonian $H_{\eff}$ of 
Eq.\ (\ref{Heff}) 
is analyzed to determine the ground-state properties 
of the localized spins as a function of conduction band 
filling $n$ and Kondo coupling strength $J > 0$. Since the 
FM double-exchange coupling ${\cal J}$ is of order $J^{2}$ 
(cf.\ Eq.\ (\ref{calJ})), it is immediate from 
Eq.\ (\ref{Heff}) that $H_{\eff}$ 
determines a FM ordering for the localized spins at stronger 
couplings $J \gg 1$ for all fillings $n < 1$. We find below that 
the FM ordering is gradually destroyed as the coupling $J$ is 
lowered. The destruction of the FM order is  
determined by the second term of Eq.\ (\ref{Heff}); 
the effective 
hamiltonian takes the form of a transverse-field Ising chain 
in the phase transition region, and the KLM undergoes a 
quantum FM-PM transition at a filling dependent critical 
coupling $J_{c}$. The critical coupling is of order unity at 
most conduction band fillings. The determination of the 
FM-PM transition, and a disussion of the 
properties of the localized spins near 
the phase boundary, are contained in subsection IIIA. In 
subsection 
IIIB, we consider the weak-coupling regime $J \ll 1$. 
At weak coupling the double-exchange ordering 
is ineffective, and $H_{\eff}$ reduces to a system of 
free localized spins in fields determined by 
conduction electron scattering (the last two terms of 
Eq.\ (\ref{Heff})). We find that $H_{\eff}$ 
determines dominant $2k_{F}$ (RKKY-like) correlations in the 
localized spins at weak coupling. Finally, in subsection IIIC, 
we plot the ground-state phase diagram for the KLM as 
determined by the effective hamiltonian.

\subsection{The FM-PM phase transition}

We begin our analysis of $H_{\eff}$
by evaluating the long-range object 
$K(j)$ in the strong-coupling FM phase.
$K(j)$, given by Eq.\ (\ref{Kj}), originates 
with the unitary transformation S. It has some similarity 
to the disorder term in the Jordan-Wigner transformation, but 
instead of counting the $S^{z}_{j'}$ only over sites to the 
left of $j$, it counts also the $S^{z}_{j'}$ to the right. 
In the thermodynamic limit $N \rightarrow \infty$, and 
using the large $j - j'$ form 
$[\phi_{\sig}(j), \theta_{\sig}(j')] = 
{\rm sign}(j - j')\, i\pi$ for the Bose field commutator,  
we may rewrite Eq.\ (\ref{Kj}) in the form 
\beqa
K(j) = \frac{Ja}{2v_{F}}\sum_{l=1}^{\infty} \epsilon_{j}(l)
\, ,
\label{Kj2}
\eeqa
where $\epsilon_{j}(l) = S^{z}_{j+l} - S^{z}_{j-l}$. 
$\epsilon_{j}(l)$ has  
possible values $0, \pm 1$ for large $l$. 
For small $j - j'$, the commutator
$[\phi_{\sig}(j), \theta_{\sig}(j')]$ grows smoothly from 
zero at $j = j'$, to  ${\rm sign}(j - j')\, i\pi$ 
at $(j - j) = {\cal O}(\alpha/a)$. The exact form of the 
commutator at short-range depends on the choice of cut-off 
function $\Lambda_{\alpha}(k)$, but all we require here  
is the general form: The effect of short-range corrections
to the Bose field commutator is just to allow  
$\epsilon_{j}(l)$ to take values between $-1$ and $1$ for 
$l \lesssim \alpha/a$. $K(j)$ is then similarly 
smoothed, and takes values between integral multiples 
of $Ja/2v_{F}$. 

By writing $K(j)$ in the form of 
Eq.\ (\ref{Kj2}), it is clear that $K(j)$ vanishes in the FM 
phase in a thermodynamically large system. 
Indeed $K(j)$ will not be appreciable until the system is 
strongly disordered. It follows that any transition out of the 
FM phase will be governed by the first two terms of the 
effective hamiltonian Eq.\ (\ref{Heff}). For convenience we 
collect these terms in the hamiltonian $H_{\crit}$ (with 
`crit' for critical):
\beqa
H_{\crit} = &-& {\cal J}\sum_{j}S^{z}_{j}S^{z}_{j+1} 
\nonumber \\
&+& A\frac{Ja}{\alpha}\sum_{j} 
\{1 + \cos(2k_{F}ja)\}S^{x}_{j} 
\label{Hcrit}
\eeqa
$H_{\crit}$ is a quantum transverse-field Ising chain, and 
a discussion of its properties occupies the remainder of this 
subsection.  A great deal is already known about the 
transverse-field Ising chain, and the following discussion is 
essentially a summary of known results as they relate to 
the KLM. Of particular importance, it is known that 
$H_{\crit}$ undergoes a quantum phase transition from a 
FM phase, to a disordered PM phase. We will outline how 
the transition is formally determined below, but before 
proceeding it is perhaps useful to consider the physics of 
the FM-PM transition in the KLM. 

$H_{\crit}$   
describes the double-exchange FM ordering being 
gradually destroyed as the coupling $J$ is lowered. 
The first term of $H_{\crit}$, which describes double-exchange, 
has been discussed extensively in subsection IIB above. The 
destruction of the double-exchange ordering may be 
understood physically as 
follows:  As the coupling $J$ is decreased the 
conduction electrons become less strongly bound to the localized 
spins, and tend to extend over spatial ranges beyond 
the effective range $\alpha$ for double-exchange ordering. 
Double-exchange becomes less effective, 
and regions of ordered localized spins begin to 
interfere as the conduction electrons extend. 
The interference leads to spin-flip processes, and are 
embodied in $H_{\crit}$ in the transverse-field (the second 
term of Eq.\ (\ref{Hcrit})). 
The transverse-field in 
$H_{\crit}$ includes two low-energy spin-flip processes by 
which the conduction electrons disorder the localized spins. 
One spin-flip process is backscattering, and is accompanied 
by a momentum transfer of $2k_{F}$ from the conduction 
electrons to the 
localized spins. Since the chain of localized spins 
will tend to order so as to reflect this transfer, the 
transverse-field corresponding to backscattering spin-flips 
is sinusoidal with modulation $2k_{F}$. The other low-energy 
spin-flip process in $H_{\crit}$ is forward scattering. 
This involves zero momentum 
transfer to the localized spins, and the 
corresponding transverse-field is a constant (i.e.\ 
has modulation zero). 

It will become clear below that either forward or backscattering 
spin-flip processes separately 
are sufficient to destroy the FM order, and bring on a FM-PM 
phase transition in the KLM. However,  
for incommensurate conduction band filling, 
the backscattering spin-flip processes introduce a 
competing periodicity in the chain of localized spins. 
It turns out that this has non-trivial consequences for certain 
properties of the localized spins near the transition. 
In the following we first indicate how the FM-PM transition 
is determined for arbitrary transverse-fields. We then 
compare the properties of the localized spins which are 
disordered 
due to forward scattering (constant transverse-field), with 
the properties in which the spin disorder is  due to 
backscattering (incommensurately modulated transverse-field). 
We point out that the special properties resulting from 
incommensurate backscattering are 
at least qualitatively reproduced 
by treating the full transverse-field in $H_{\crit}$ as 
a random variable with the appropriate (displaced cosine) 
distribution. The subsection concludes with a summary of the 
properties of the random transverse-field Ising chain as they 
relate to the transition region in the KLM.

\subsubsection*{Determination of the phase transition}

Using a Jordan-Wigner transformation to spinless 
fermions $a^{}_{j}, a^{\dg}_{j}$, and in the 
thermodynamic limit, $H_{\crit}$ may be 
written 
\beqa
H_{\crit} = \sum_{j,l}\left\{ a^{\dg}_{j}A_{jl}a^{}_{l} 
+ \frac{1}{2}\left( a^{\dg}_{j}B_{jl}a^{\dg}_{l} + 
{\rm h.c.}\right) \right\} 
\label{matrixform}
\eeqa
to an additive constant, where $(A_{jl})$ and 
$(B_{jl})$ are real 
symmetric and antisymmetric matrices, respectively, 
with non-zero entries 
\beqa
A_{jj} &=& h_{j} \equiv A\frac{Ja}{\alpha}
\{1 + \cos(2k_{F}ja)\} 
\nonumber \\
A_{jj+1} &=& A_{j+1j} = B_{jj+1} = -B_{j+1j} 
= -{\cal J}/4\, .
\nonumber 
\eeqa
The quadratic form Eq.\ (\ref{matrixform}) may be 
diagonalized for any transverse-field $h_{j}$ by 
using the method of Lieb {\it et al.}
\cite{Lieb}. This gives 
\beqa
H_{\crit} = \sum_{k} \omega_{k} \eta^{\dg}_{k}
\eta^{}_{k} 
\label{Hcritdiag}
\eeqa
to an additive constant, where $\eta^{\dg}_{k}, 
\eta^{}_{k}$ are creation and annihilation operators for 
free spinless fermions, and 
where the energies $\omega^{2}_{k}$ are eigenvalues 
of the symmetric matrix $(A + B)(A - B)$. As the 
coupling $J$ is decreased, $H_{\crit}$ undergoes 
a quantum order-disorder transition from a 
FM phase, to a quantum disordered PM phase, 
signalled by the breakdown of long-range 
correlations between the localized spins, 
and a continuously vanishing spontaneous 
magnetization. The critical line for the transition
is determined by the critical coupling $J_{c}$ which 
solves \cite{Pfeuty79}
\beqa
{\cal J}^{N} - 2^{N}\prod_{j=1}^{N}h_{j} = 0 
\label{implicit}
\eeqa
as $N \rightarrow \infty$. The free energy of the localized 
spins becomes non-analytic 
at points satisfying Eq.\ (\ref{implicit}). 

The FM-PM transition at the coupling $J_{c}$ is generic to 
transverse-field Ising chains, and does not assume a 
particular form for the transverse-field $h_{j}$ 
\cite{Pfeuty79}. For example, 
if we consider only forward scattering spin-flip processes 
in the KLM, the transverse-field $h_{j} = AJa/\alpha$ is a 
constant. Solving Eq.\ (\ref{implicit}) with 
$h_{j} = AJa/\alpha$ gives the quantum 
critical line for the FM-PM transition at
\beqa
J_{c} = 
\frac{8\pi^{2}A\sin(\pi n /2)}{\alpha \int_{0}^{\infty}
dk\, \cos(ka)\Lambda^{2}_{\alpha}(k)}\, .
\label{Jc}
\eeqa
A detailed discussion of the properties of the Ising chain with 
a constant transverse-field, which describes the FM-PM 
transition in the KLM with backscattering neglected, is given 
by Pfeuty \cite{Pfeuty70}. 
As a second example, if we consider only backscattering 
spin-flip processes in the KLM, the transverse-field 
$h_{j} = AJa \cos(2k_{F}ja)/\alpha$ is sinusoidal. In real 
heavy-fermion materials, the number of available conduction 
electrons per localized spin will in general be irrational 
\cite{Strong}. In this case the transverse-field due to 
backscattering has an incommensurate modulation $2k_{F}$ 
with respect to the underlying lattice of localized spins. 
Nonetheless a FM-PM transition still occurs. 
As shown in Ref.\ [30], the solution of 
Eq.\ (\ref{implicit}) for incommensurately modulated 
transverse-fields yields a coupling $J_{c}$ as in 
Eq.\ (\ref{Jc}) for the constant transverse-field. Thus the 
critical line for the FM-PM transition in the KLM with only
backscattering spin-flip processes coincides with the critical 
line for the transition 
in the KLM with only forward scattering.

\subsubsection*{Effects of the form of the transverse-field} 

While the FM-PM transition itself is largely independent 
of the details of the transverse-field,  
there are significant differences in the properties of the 
localized spins on either side of the transition depending on 
the particular form of the transverse-field. 
The differences are most clearly apparent in the wavefunctions 
corresponding to the free fermions $\eta_{k}$ of the 
diagonalized hamiltonian Eq.\ (\ref{Hcritdiag}). 
The wavefunctions are 
always extended, or Bloch-like, for a constant transverse-field
\cite{Pfeuty70}.
For an incommensurately modulated transverse-field, the behavior 
of the wavefunctions is far more complex. The Ising chain 
with an incommensurate transverse-field has been studied 
extensively by Satija {\it et al}. 
\cite{Satija89,Satija90,Satija94}. The model is 
important as it has localized states in 1D, and thus 
provides a link to random systems 
\cite{Satija94}. Numerical studies \cite{Satija89} show 
that the wavefunctions corresponding to the free fermions 
$\eta_{k}$ are localized in the disordered PM phase, and undergo 
a spectral transition at the FM-PM phase boundary.
In the FM phase, the wavefunctions are 
self-similar and the eigenvalue spectrum forms a 
Cantor set. Since the correlation functions for the localized 
spins are determined by the wavefunctions, the KLM with 
backscattering possesses far different properties 
to the KLM with only forward scattering spin-flip processes, 
even though both undergo a FM-PM transition. Differences in 
the eigenvalue spectrum $\omega_{k}$ of the diagonalized 
hamiltonian Eq.\ (\ref{Hcritdiag}) lead to a similar 
conclusion regarding thermodynamic properties; since the KLM 
with incommensurate backscattering has a fractal eigenvalue 
spectrum, its thermodynamics are far different to the KLM 
with forward scattering, which has the more standard 
(cosine-type) eigenvalue spectrum \cite{Pfeuty70}. 

The situation becomes yet more complex when we consider all 
possible low-energy spin-flip processes available to 
the conduction electrons. $H_{\crit}$ includes forward 
scattering with zero momentum transfer, and is represented by a 
constant transverse-field. $H_{\crit}$ also includes 
backscattering with an incommensurate momentum transfer 
$2k_{F}$, and is represented by a $2k_{F}$ 
sinusoidal transverse-field. $H_{\crit}$ does not include 
spin-flip interactions with momentum transfers at higher 
harmonics of $2k_{F}$: at $4k_{F}$, $6k_{F}$, and so on. The 
higher harmonics will arise in a bosonization treatment which 
includes non-linear corrections to the conduction electron 
dispersion relation \cite{Voit}. These corrections are 
very weak compared with the forward 
and backscattering spin-flip processes, and it is usual to 
neglect them. However, the addition 
of (even weak) higher harmonics in $2k_{F}$ to the  
transverse-field $h_{j}$ will greatly alter the solution of Eq.\ 
(\ref{implicit}). Instead of one solution, there now occur 
an infinite number of solutions to Eq.\ (\ref{implicit}), 
and these occupy a finite region of the parameter 
space \cite{Satija89}. The series of solutions is 
reflected in numerical studies 
on the Ising chain with a transverse-field containing 
more than one incommensurate harmonic \cite{Satija89,Satija90}. 
The region of spectral transitions becomes broadened, and 
the wavefunctions 
corresponding to the free fermions $\eta_{k}$ of Eq.\ 
(\ref{Hcritdiag}) are observed to undergo a cascade of 
transitions between extended, critical, and 
localized behavior. The transitions in the wavefunctions occupy 
a finite region of the parameter space, 
and coincide with the solutions of Eq.\ (\ref{implicit}) 
at which the free energy becomes  
non-analytic. While the region of spectral 
transitions becomes broadened, this is not the case for the 
magnetic transition. The FM-PM transition, signalled 
by the vanishing of longe-range correlations between the 
localized spins, is observed to remain sharp \cite{Satija90}.  

The behavior observed by Satija {\it et al.} in the 
numerical studies 
discussed above is qualitatively identical to the 
behavior of the Ising chain with a random transverse-field. 
(Properties of the random transverse-field Ising chain are 
disussed extensively by Fisher \cite{Fisher}, and are 
summarized below.) To see this identification, 
note that the central 
feature of the random transverse-field Ising chain is that 
dilute regions of FM order may survive into the PM phase, and 
similarly that dilute regions of disorder may continue into 
the FM phase. This feature is at the heart of Fisher's results
\cite{Fisher}, and is shared by $H_{\crit}$ for 
incommensurate $k_{F}$: As discussed above, a broadened region 
of spectral transitions about the true FM-PM transition occurs 
in the KLM with incommensurate conduction band filling. Thus 
there are small regions in the PM phase in which 
the localized spins exhibit behavior normally associated with 
the FM phase, and vice versa. To further pursue the 
identification between a random transverse-field, and 
that present in $H_{\crit}$ for the KLM, recall that the 
spectral transitions occur at points satisfying Eq.\ 
(\ref{implicit}) at which the free energy becomes non-analytic. 
There is an immediate identification between these non-analytic 
points, and the Griffiths singularities 
\cite{Griffiths} present in random models, in which 
thermodynamic quantities such as the magnetization become 
singular in a range of parameter space about the non-random 
transition. (See Ref.\ [33] for the Griffiths regions in the 
random transverse-field Ising chain.) 

The behavior of $H_{\crit}$ for incommensurate conduction band 
filling admits of a natural physical interpretation. The 
conduction band does not share the periodicity of the lattice 
of localized spins, 
and is unable to either totally order or totally disorder the 
lattice as the FM-PM transition is crossed. There remain dilute 
regions of double-exchange ordered 
localized spins into the PM phase as 
only a quasi-commensurate fraction of the conduction electrons 
become weakly-bound, and become free to scatter along the chain, 
at the FM-PM transition. The remaining ordered regions are 
dilute enough that no long range correlations remain, but 
their existence dominates the low-energy properties of the 
localized spins near the transition.

These considerations lead us to treat the transverse-field 
$h_{j} = AJa\{1 + \cos(2k_{F}ja)\}/\alpha$ of $H_{\crit}$ 
as a random variable, so that $h_{j}$ is chosen from the 
displaced cosine distribution $\rho(h)dh$ where
\beqa
\rho(h) = \frac{\alpha}{\pi AJa}
\frac{1}{\sqrt{1- (\alpha h/AJa - 1)^{2}}}\, . 
\label{rhoh}
\eeqa
As discussed above, this treatment of $h_{j}$ does not 
alter the basic FM-PM transition described by $H_{\crit}$, 
and thus is not needed in order to plot the phase diagrams 
of the KLM in Figs.\ 3 and 6 below. However, it does 
account for the properties of the localized spins near the 
FM-PM transition as observed by Satija {\it et al.} in their 
numerical simulations. 
We conclude by noting that at low conduction band filling 
the treatment of 
$h_{j}$ in Eq.\ (\ref{rhoh}) follows an analogous treatment 
in spin glass systems (cf.\ Ref.\ [3]).

\subsubsection*{Properties of the localized spins near 
criticality}

Results on the random transverse-field Ising chain 
may be obtained from Fisher \cite{Fisher}, who uses 
an approximate real-space renormalization-group (RG)
analysis, which nonetheless yields 
asymptotically exact results at low-temperatures
near criticality. Following Ref.\ [33], the 
critical coupling for the KLM is given by
\beqa
J_{c} = 
\frac{4\pi^{2}A\sin(\pi n /2)}{\alpha \int_{0}^{\infty}
dk\, \cos(ka)\Lambda^{2}_{\alpha}(k)}\, .
\label{Jctot}
\eeqa
The critical line thus retains the form 
of Eq.\ (\ref{Jc}) for forward or 
backscattering separately, but is down by a factor of 2 as 
both spin-flip disorder processes are included. 
It is convenient to measure deviations from criticality by 
\cite{Fisher}
\beqa
\delta = \left[ {\rm var}(\log h) \right]^{-1} 
\log \left\{ 
\frac{4\pi^{2} A 
\sin(\pi n/2)}{J\alpha\int_{0}^{\infty}dk\, 
\cos(ka)\Lambda_{\alpha}^{2}(k)} \right\} ,
\label{delta}
\eeqa
where the measure of randomness is 
\beqa
{\rm var}(\log h) = \sum_{n=1}^{\infty} 
\left\{\frac{1}{n}
\frac{1.3. \cdots .(2n-1)}{2.4. \cdots .(2n)}
\sum_{m=1}^{2n-1}\frac{1}{m} \right\} 
- \log^{2}2\, . 
\nonumber 
\eeqa
$\delta = 0$ on the critical line, is positive 
in the disordered PM phase, and negative in the 
FM phase. 

The distinctive feature of Fisher's RG  
analysis is that it focuses on anomalous clusters of 
double-exchange ordered localized spins  
which survive for small $\delta$ into the PM 
phase, and similarly, rare disordered regions 
close to criticality in the FM phase. These are due 
to the incommensurability of the conduction band 
filling with respect to the lattice of localized spins, 
and the consequent inability of the conduction 
band, as a single many-body entity, to either 
totally order or totally disorder the localized spins 
as the transition is crossed. It is the anomalous 
ordered (disordered) regions of localized spins in 
the PM (FM) phase which are responsible for the
Griffiths singularities. Although these anomalous 
regions are very dilute, they dominate the 
low-energy properties of the spin chain. Thus, while 
typical correlations are much as in the 
constant transverse-field Ising chain, the 
measurable mean correlations are dominated by 
the anomalous regions, and consequently greatly alter the  
low-energy behavior.

An important prediction of our theory 
of the phase transition in the KLM 
is that the spontaneous magnetization grows 
continuously from criticality into the FM phase: 
For the random transverse-field \cite{Fisher} 
\beqa
M_{0}(\delta) \sim (-\delta)^{\beta}\, ,
\quad \quad \delta < 0\, ,
\label{Mo}
\eeqa
where $\beta = (3 - \sqrt{5})/2 \approx 0.38$ 
\cite{caveat1}. This disagrees with numerical 
diagonalization results on small systems 
\cite{Tsunetsugu}, which see a discontinuous jump 
in $M_{0}$ at least at larger fillings, 
but note that regions 
of intermediate $M_{0}$ have been observed in  
related studies \cite{Moukouri96}, and in  
small systems at lower fillings \cite{Tsunetsugu}. 
Indeed a discontinuous jump in 
$M_{0}$ immediately above the transition 
seems difficult 
to understand in a thermodynamically large system, 
given that the ordering is due to 
double-exchange, and that the electron 
spin degrees of freedom are not frozen until deep 
into the FM phase \cite{Moukouri95,caveat2}. 

Using Ref.\ [33], we summarize the properties 
of the 1D KLM which are relevant to the transition  
region of small $\delta$. The 
mean spin-spin correlation function is defined by 
\beqa
{\overline C}(x) = 
{\overline{\langle S^{z}_{j}S^{z}_{j+x}\rangle}}\, ,
\nonumber
\eeqa
where the average is over $\rho(h)$, and where for 
convenience $x$ denotes a continuous and 
positive variable. ${\overline C}(x)$ is dominated  
by atypically large correlations and for small 
$|\delta|$ in the FM phase decays as   
\beqa
{\overline C}(x) \sim  
M_{0}^{2}(\delta) + {\rm const.}|\delta|^{2\beta}
(\xi/x)^{5/6}e^{-3(\pi x/\xi)^{1/3}}e^{-x/\xi} 
\nonumber 
\eeqa
for $x \gg \xi$. The 
correlation length $\xi \approx 1/\delta^{2}$. (For 
typical pairs of spins the correlation length 
$\xi \approx \delta^{-1}$; the exponent is the same as in 
the Ising chain with a constant transverse-field.) 
At criticality, $\delta = 0$, the decay 
is power law: ${\overline C}(x) \sim 
x^{-\beta}$ as $x \rightarrow \infty$. For small 
$\delta$ in the PM phase,  
\beqa
{\overline C}(x) \sim
\delta^{2\beta}
(\xi/x)^{5/6}e^{-(3/2)(2\pi^{2} x/\xi)^{1/3}}e^{-x/\xi} 
\label{Cx}
\eeqa
where $x \gg \xi \approx 1/\delta^{2}$. Note that 
${\overline C}(x)$ decays more rapidly to 
$M_{0}^{2}(\delta)$ in the FM phase, than it decays to 
zero in the PM phase. 

At low temperatures $T$ close to the 
transition, ${\overline C}(x,T)$ decays exponentially at large 
distances with a correlation length $\xi_{T}$. In the 
FM phase, $\xi_{T}$ diverges as a continuously variable 
power law of $T$:
\beqa
\xi_{T} \approx 
e^{2\Gamma_{T}|\delta|}/4\delta^{2} \quad \quad
\Gamma_{T}|\delta| \rightarrow \infty  
\nonumber 
\eeqa
in the FM phase, where $\Gamma_{T}$ is a characteristic 
scale, given by $\Gamma_{T} = 
\log( {\rm max}\{{\cal J}, h_{j}\}/T)$ 
at fixed $J$ and $n$ close to the transition. 
At criticality the correlation length is 
$\xi_{T} \approx 4\Gamma^{2}_{T}/\pi^{2}$, while in the 
PM phase 
\beqa
\xi_{T} \approx (\delta^{2} + \pi/\Gamma_{T}^{2})^{-1}  
\quad \quad \Gamma_{T}\delta \gg 1\, . 
\nonumber 
\eeqa
The correlation lengths $\xi_{H}$ for the long-range 
exponential decay of the correlations 
${\overline C}(x, H)$ in small 
applied fields $H$ along $z$ have 
identical functional forms to those of $\xi_{T}$ above. We 
note only that in the FM phase, $\xi_{H} \sim 
H^{-2|\delta|}$ as $H \rightarrow 0$. This reflects the 
development of long-range order, and shows a power 
law dependence on $H$. (See Fisher \cite{Fisher} 
for more details).

The magnetization in small positive applied fields 
$H$ along the $z$ direction is obtained \cite{Fisher} 
using an exact critical scaling function. Close to 
the critical line in the FM phase this gives
\beqa
M(\delta ,H) \sim 
M_{0}(\delta)[1 + {\cal O}(\delta H^{2|\delta |}\log H)] 
\nonumber
\eeqa
at zero temperature. At criticality, 
$M(\delta, H) \sim |\log H |^{-\beta}$, and in the PM phase 
\beqa
M(\delta, H) \sim
\delta^{1+\beta} H^{2\delta} |\log H |  \, .  
\label{M} 
\eeqa
Close to the transition in both phases the magnetization 
is highly singular. In the PM phase the magnetization has a 
power law singularity with a continuously variable exponent 
$2\delta$, and the linear susceptibility is 
infinite for a range of $\delta$ into the PM 
phase. The susceptibility remains infinite 
(with a continuously variable exponent) close to the 
transition into the FM phase. The low temperature 
linear susceptibility $\chi(T)$ takes the form 
\beqa
\chi (T) \sim 
T^{2\delta -1}(-\delta)^{-2(1-\beta)} 
\nonumber 
\eeqa
in the FM phase, and $T \chi (T)$ diverges as $T \rightarrow 0$. 
Note that the latter property was conjectured 
by Troyer and W\"{u}rtz on the basis of their quantum 
Monte Carlo results \cite{Troyer}.
At criticality, $\chi (T) \sim 
T^{-1} |\log T|^{2(1-\beta)}$, and close to the transition 
in the PM phase, 
\beqa
\chi (T) \sim 
\delta^{-4(1-\beta)}T^{2\delta - 1}(\log T)^{2}. 
\label{chiT}
\eeqa
$T \chi (T)$ vanishes as $T \rightarrow 0$ in the PM 
phase. The zero-field 
specific heat at low temperatures close to the 
transition is given by 
\begin{eqnarray}
C_{v}(T) \sim |\delta|^{3} 
T^{2|\delta|}[1 + {\cal O}(T)^{2|\delta|}] 
\nonumber
\end{eqnarray}
in either the FM or PM phases. At criticality, the specific 
heat $C_{v}(T) \sim | \log T|^{-3}$.

\subsection{Weak-coupling}

Well below the transition, where $Ja/v_{F}$ is small 
and the FM interaction Eq.\ (\ref{FMterm}) is 
negligible, the ordering of the localized spins is 
governed by the last two terms of 
the effective hamiltonian Eq.\ (\ref{Heff}). 
To determine the dominant correlations in this 
strongly disordered phase, 
it suffices to take eigenvalues for 
$\epsilon_{l}(j)$ in the 
long-range object $K(j)$ (cf.\ Eq.\ (\ref{Kj2})). 
$K(j)$ then fluctuates about zero incoherently, 
depending on the global $S^{z}_{j}$ configuration.
The effective hamiltonian corresponds to free 
localized spins in  
$x$ and $z$ fields determined by 
conduction electron scattering. 
The free localized spin problem is straightforwardly 
diagonalized by standard methods \cite{Wagner}, and 
yields a ground-state $S^{z}_{j}$ configuration 
$|\psi_{0}\rangle$ given by
\beqa 
\exp  \left\{ 
i\sum_{j} \tan^{-1}\left(
\frac{\cos[K(j)] 
+ \cos(2k_{F}ja)}{\sin[K(j)]\sin(2k_{F}ja)} \right)
S^{y}_{j}\right\} |\downarrow\rangle \, ,
\nonumber
\eeqa
where $|\downarrow \rangle$ is the state with 
$S^{z}_{j} = -1/2$ for all $j$. The 
dominant $2k_{F}$ modulations in 
$|\psi_{0} \rangle$ are manifest, 
and are superimposed on an incoherent backgound: 
\beqa
\langle\psi_{0}| S^{z}_{j} S^{z}_{j+x} 
|\psi_{0}\rangle 
\approx \sin[2k_{F}ja] \sin[2k_{F}(ja+x)] 
\nonumber
\eeqa
to an incoherent normalization.
This is observed at weak-coupling 
in numerical simulations 
\cite{Troyer,Moukouri95,Caprara}, and is 
called the RKKY regime. (Note, however, that the RKKY 
interaction strictly diverges in 1D, and there  
is no lower bound on the ground-state energy for 
the RKKY hamiltonian, even for arbitrarily small  
$J$ \cite{strong-coupling}.  The divergence is  
typical of perturbation expansions in 1D, and does 
not occur in higher dimensions.)

\subsection{Phase diagram}

The behavior identified in the previous subsections 
is in complete qualitative agreement with the results 
of numerical simulations on larger systems 
\cite{Troyer,Moukouri95,Caprara}. 
To establish quantitative agreement,   
i.e.\ to plot the critical line,
we are presented with two obstacles. 
The critical line Eq.\ (\ref{Jctot}) with both forward and 
backscattering spin-flip interactions included may be  
written
\beqa
J_{c}a = 
\frac{2\pi^{2}Av_{F}}{\alpha \int_{0}^{\infty}
dk\, \cos(ka)\Lambda^{2}_{\alpha}(k)}\, .
\label{fullJc}
\eeqa
The first obstacle in using Eq.\ (\ref{fullJc}) is
somewhat trivial, and relates 
to the global scaling of the critical line: The number $A$ 
comes from the normalization of the Bose representations 
for spin-flip and backscattering electron interactions. 
It depends significantly on the cut-off function 
$\Lambda_{\alpha}(k)$, and moreover relates to the 
normalization of Bose representations only in the limit 
of long wavelengths (cf.\ Eq.\ (\ref{Crjsigbose})). 
The second obstacle  
is the dependence $\alpha = \alpha(n,J)$ which measures 
the effective range of the double-exchange interaction.  
This is a non-trivial quantity 
in a thermodynamically large system. 
In our previous work \cite{Honner}, we made the 
approximation of neglecting any functional dependence of 
$\alpha$ on $n$ and $J$, and determined $A$ by a fit to 
numerically determined points. The resulting phase diagram 
\cite{Honner} correctly gives the general ground-state 
phase diagram of the KLM for $J > 0$, and indicates 
schematically the regions where Griffiths singularities occur 
for incommensurate filling, together with the crossover 
to the strongly-disordered 
(RKKY-like) regime at weak-coupling. Here we provide a 
more detailed analysis, and use numerically determined 
phase transition points to determine the functional 
dependence of $\alpha$ on the critical line. Note that 
contrary to all previous results, the results of this 
subsection rely crucially on numerics. 

We estimate first the constant $A$. From bosonization, we  
know that $\alpha \gtrsim {\cal O}(k_{F})^{-1}$, 
and so $\alpha$ will diverge as the filling $n \rightarrow 0$.
From Eq.\ (\ref{fullJc}) it follows that 
\beqa
J_{c} \rightarrow 2\pi^{3}A\, n\, , \quad \,
{\rm as}\,\,\,  n \rightarrow 0\, .
\label{smalln}
\eeqa
Note the agreement with the 
exact solution of Sigrist {\it et al.} 
\cite{one-electron} for the 
KLM with one conduction electron; the system is FM for all 
finite $J$. Recall also that the exact solution gives 
$\alpha/a = \sqrt{2/J}$ for small $J$. It follows that $\alpha$ 
diverges at criticality in agreement with the result 
from bosonization as $n \rightarrow 0$.  
Using numerical results for $J_{c}$ 
at the smallest available filling ($J_{c} = 0.455$ 
at $n = 0.2$ from the infinite-size DMRG simulation 
of Caprara and Rosengren \cite{Caprara}), we conclude 
from Eq.\ (\ref{smalln}) that $2\pi^{2}A \approx 0.7$.  

\begin{figure}[tb]
\centering
\epsfig{file=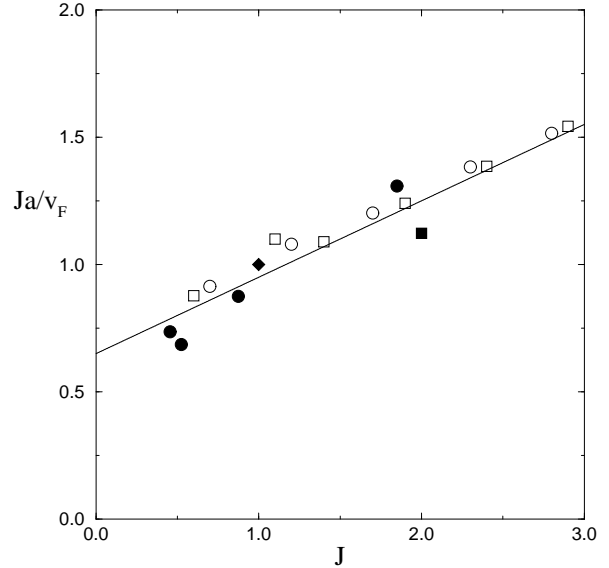,height=9cm}
\caption{Plot of the dimensionless parameter $Ja/v_{F}$, 
which characterizes double-exchange FM, against 
coupling $J > 0$ for numerically determined 
FM-PM transition points $J_{c}$: 
the filled diamond is the quantum Monte Carlo
result for systems up to 24 sites from Ref.\ [4]; open 
circles and squares are exact numerical diagonalization 
results for the 8 and 9 site chain, respectively, 
from Ref.\ [5]; the filled square is the DMRG
result for systems up to 75 sites from Ref.\ [6];  
the filled circles are infinite-size DMRG results 
from Ref.\ [7]. $J_{c}a/v_{F} \approx 0.7$ for vanishing $J$. 
The straight 
line of best fit is given, and shows good agreement 
with the spread of numerical results, together with 
the expected result as $J \rightarrow 0$.} 
\end{figure}

We now account for the functional dependence of $\alpha$, and 
plot the critical line. $\alpha$
enters the critical line equation in the  
denominator of the right hand side of 
Eq.\ (\ref{fullJc}). This factor may be determined 
independently of a choice for the cut-off function 
$\Lambda_{\alpha}(k)$ by using numerically determined 
FM-PM transition points. 
In Fig.\ 2 
we plot the dimensionless parameter $Ja/v_{F}$ against 
$J$ for available numerical data
\cite{Troyer,Tsunetsugu,Moukouri95,Caprara}.  $Ja/v_{F}$ 
characterizes double-exchange in our theory, and gives 
the denominator of Eq.\ (\ref{fullJc}) at criticality. 
The functional dependence is linear, and we give 
in Fig.\ 2 the straight line of best fit. This line 
gives $2\pi^{2}A = 0.65$ as $n \rightarrow 0$, in 
agreement with the estimate $\approx 0.7$ from $n = 0.2$
given in the previous paragraph. It is reasonable to conclude that the 
deviations in the numerically determined points for 
$Ja/v_{F}$ from the straight line are reflections 
of the different critical values determined in 
different simulations. The line of Fig.\ 2, 
together with Eq.\ (\ref{fullJc}),
determines the critical line at
\beqa
J_{c} = \frac{1.3 \sin(\pi n/2)}{1-0.6\sin(\pi n/2)}\, , 
\quad  \quad J > 0\, .
\label{Jcafm} 
\eeqa
The resulting phase diagram is given in Fig.\ 3. 

\begin{figure}[tb]
\centering
\epsfig{file=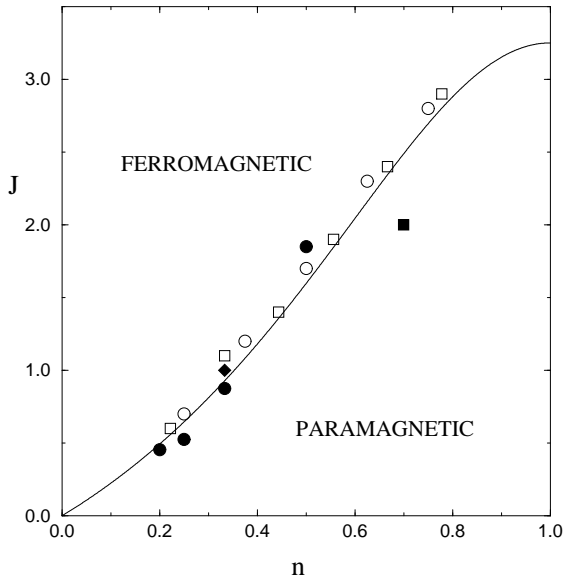,height=9cm}
\caption{Ground-state phase diagram of the 1D KLM with 
$J > 0$. The critical line is from 
Eq.\ (\ref{Jcafm}), and uses the line of Fig.\ 2. 
Numerically determined critical points are as in 
Fig.\ 2. At incommensurate fillings, 
there are Griffiths singularities in the free 
energy in a finite region of the parameter space 
about the critical line. At small $Ja/v_{F}$ in 
the paramagnetic phase, the 
system presents an RKKY-like behavior with dominant 
correlations in the localized spins at $2k_{F}$ 
of the conduction band.}
\end{figure}

The line of Fig.\ 2 determines the effective range 
$\alpha$ of the double-exchange interaction on the 
transition line. Choosing the exponential cut-off 
function $\Lambda_{\alpha}(k) = e^{-\alpha|k|/2}$ for 
simplicity, we have 
\beqa
\frac{1}{\alpha\int_{0}^{\infty}dk\, 
\cos(ka) \Lambda_{\alpha}^{2}(k)} = 
1 + (a/\alpha)^{2}\, .
\label{alphaexp}
\eeqa
For this choice of cut-off function, the line of 
Fig.\ 2 gives $\alpha/a =
\sqrt{2.1/J}$ at the transition. This compares with 
the result $\sqrt{2/J}$ obtained in the exact solution 
of the KLM with one conduction electron \cite{one-electron} 
just above the critical point at vanishing $J$. 
The filling dependence of  
$\alpha$ may be determined by using Eq.\ (\ref{Jcafm}) to 
write $\alpha/a = \sqrt{2.1/J}$ in terms of $n$. The result 
is plotted in Fig.\ 4.

\begin{figure}[tb]
\centering
\epsfig{file=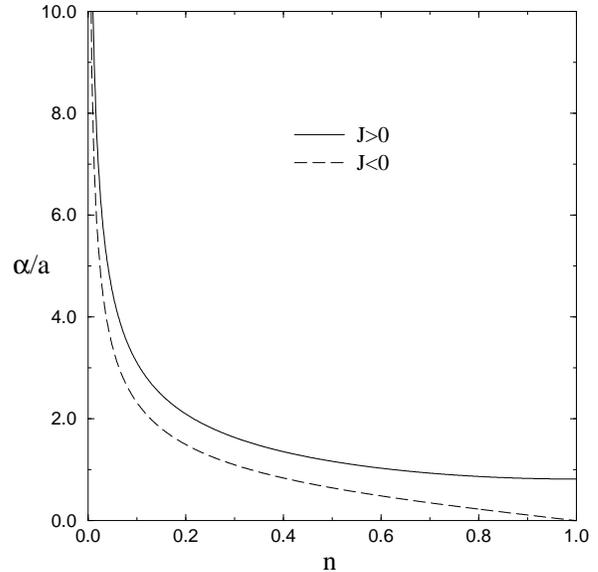,height=9cm}
\caption{The effective range $\alpha$ of the double-exchange 
interaction in units of the lattice spacing against 
filling $n$ on the critical line. An 
exponential cut-off function $\Lambda_{\alpha}(k) 
= \exp -(\alpha|k|/2)$ has been chosen. The vanishing 
of the range at half-filling in the $J < 0$ KLM 
leads to a divergence in the critical line as 
half filling is approached.}
\end{figure}

\section{THE FM-PM TRANSITION IN RELATED MODELS}

In this section we consider some variants of the 
1D KLM of Eq.\ (\ref{klm}), and use our methods 
to describe the FM-PM transition at partial conduction 
band filling in these models. We consider firstly the 
effects of interactions between the  
electrons, and show that for repulsive 
interactions the phase boundary is always pushed to 
lower $J$ values, in agreement with available exact 
and numerical results \cite{Yanagisawa94,Moukouri96}. 
Second, we consider the KLM with a FM
$J < 0$ coupling. Our 
method extends to this model, and predicts the same 
class of phase transition as for $J > 0$. 
Using available numerical results \cite{Yunoki}, we 
follow the analysis of Sec.\ III to determine the 
critical line and give the resulting phase diagram.

\subsection{Interacting conduction band} 

To determine the effects on the ordering of the 
localized spins due to 
interactions between the electrons, consider adding 
to the standard KLM hamiltonian of Eq.\ (\ref{klm}) 
the Hubbard interaction term 
\beqa
V = U\sum_{j} n_{j\uparrow}n_{j\downarrow}\,  .  
\nonumber
\eeqa 
In terms of density operators, the interaction may be written
\beqa 
V &=& \frac{U}{4N}\sum_{k} 
  [(\rho_{+}(k) + \rho_{-}(k))(\rho_{+}(-k) + \rho_{-}(-k))
\nonumber \\ 
  & & \quad
  -(\sig_{+}(k) + \sig_{-}(k))(\sig_{+}(-k) + \sig_{-}(-k))],
\nonumber
\eeqa
where the charge and spin density operators are 
defined in Eq.\ (\ref{rhosig}). 
It will be sufficient for our purposes to consider only 
forward scattering contributions to $V$. (For weak 
repulsive interactions, $U$ small and positive,  
backscattering interactions renormalize to zero 
\cite{Solyom}.) Within the Bose description, this is equivalent 
to attaching the weight $\Lambda_{\alpha}(k)$ to the 
density fluctuations in $V$. The interaction 
then reduces to standard Tomonaga-Luttinger--type, with 
forward scattering interactions described by bosonic 
density operators. 
The pure conduction band part $H_{0} 
+ V$ of the interacting KLM may now be straightforwardly 
diagonalized via a Bogoliubov transformation 
$\exp({\rm S}_{B})$ 
where ${\rm S}_{B} = \sum_{\nu = \rho,\sig}
{\rm S}_{\nu}$ and 
\beqa
{\rm S}_{\nu} & = &
\frac{\pi}{2L}\log\left(\frac{v_{\nu}}{v_{F}}\right) 
\nonumber \\
& \times &\sum_{k>0}\frac{1}{k}[\nu_{+}(k)\nu_{-}(-k) - 
\nu{-}(k)\nu_{+}(-k)]\Lambda_{\alpha}^{2}(k).
\nonumber
\eeqa
The charge and spin velocities are 
\beqa
v_{\rho} &=& v_{F} \sqrt{1 + Ua/\pi v_{F}}\,\, ,
\nonumber \\
v_{\sig} &=& v_{F} \sqrt{1 - Ua/\pi v_{F}}\, \, .      
\nonumber 
\eeqa
By comparison with the exact Bethe ansatz solution, 
these velocities are correct to leading order in 
$U$. Corrections to the velocities at stronger 
couplings are given by Schulz \cite{Schulz}. (Note that the 
spin velocity $v_{\sig}$ does not go complex as $U$ 
increases, but smoothly goes to zero.) 
Under ${\rm S}_{B}$ the 
Bose fields transform as 
\beqa
\tilde{\phi}_{\nu}(j) &=& \sqrt{v_{F}/v_{\nu}}\, 
\phi_{\nu}(j),
\nonumber \\
\tilde{\theta}_{\nu}(j) &=& \sqrt{v_{\nu}/v_{F}}\, 
\theta_{\nu}(j),
\label{tildebose} 
\eeqa
while
\beqa
\tilde{H}_{0} &+& \tilde{V} 
= \sum_{\nu = \rho ,\sig}H_{\nu}, 
\nonumber \\
H_{\nu} &=& \frac{v_{\nu}a}{4\pi}\sum_{j} 
\left\{ \Pi_{\nu}^{2}(j) + [\partial_{x} 
\phi_{\nu}(j)]^{2}\right\} 
\nonumber
\eeqa
to an additive constant. Under  
${\rm S}_{B}$, 
the bosonized KLM with interactions between 
the electrons now takes the same basic 
form as the original bosonized hamiltonian of 
Eq.\ (\ref{bklm}). The only differences are that 
the first term in Eq.\ (\ref{bklm}) is replaced 
by $\sum_{\nu}H_{\nu}$, and that the Bose fields 
in the remaining terms are replaced by their scaled 
forms as in Eq.\ (\ref{tildebose}). Proceeding much 
as in the original $U = 0$ problem, we choose the 
transformation $\exp({\rm S})$ where
\beqa
{\rm S} = i\frac{Ja}{2\pi}\sqrt{\frac{v_{F}}{v_{\sig}^{3}}} 
\sum_{j} \theta_{\sig}(j)\,S^{z}_{fj}, 
\label{unit}
\eeqa
and obtain a transformed hamiltonian similar to 
Eq.\ (\ref{tbklm}). The important difference is 
that the prefactor of the double-exchange FM term of 
Eq.\ (\ref{FMterm}) is increased: 
\beqa
\frac{J^{2}a^{2}}{4\pi^{2}v_{F}} \rightarrow
\frac{J^{2}a^{2}}{4\pi^{2}v_{F}}\,
\frac{1}{1 - Ua/\pi v_{F}}\,.
\label{52}
\eeqa
Following exactly the analysis of Sec.\ II, we obtain  
an effective hamiltonian for the localized spins. This 
determines a quantum order-disorder transition between 
FM and quantum disordered PM phases, with a 
critical coupling $J_{c}(U)$ which is down from the 
$U = 0$ critical coupling by a factor 
$1 - Ua/\pi v_{F}$. For stronger interactions between 
the conduction electrons, the spin velocity of 
Schulz \cite{Schulz} should be used in the 
transformation Eq.\ (\ref{unit}). 

The effect of a repulsive Hubbard interaction between 
the electrons is then as follows. Double-exchange 
is characterized by the enhanced dimensionless 
constant $Ja\sqrt{v_{F}/v_{\sig}^{3}}$. The FM phase 
becomes more robust, and the FM-PM phase boundary is 
pushed to lower values of $J$. We expect 
this on physical grounds: The repulsion between the 
electrons tends to keep the regions of double-exchange 
ordered localized spins from interfering. Spin-flip disorder 
processes are thereby 
reduced. Our result is consistent 
with the numerical work of Moukouri {\it et al.} 
\cite{Moukouri96} on the KLM with $t$-$J$ 
interacting electrons; reduced critical 
couplings $J_{c} \approx 0.8, 1, 1.2$ are 
determined for fillings $n = 0.5, 0.7, 0.9$, 
respectively. Moreover, since 
$v_{\sig} \rightarrow 0$ 
as $U \rightarrow \infty$ \cite{Schulz}, so  
that $J_{c} \rightarrow 0$, our result coincides 
with the rigorous result of Yanagisawa and Harigaya 
\cite{Yanagisawa94} for infinite repulsive 
electron interactions.

\subsection{The KLM with a FM coupling}

The KLM with a FM coupling $J < 0$ 
is an effective model for CMR materials 
\cite{Zang}. The localized spins in this model 
are three $t_{2g}$ Mn $d$-electrons, and  
have spin 3/2. The properties of the 
model of interest here are largely independent  
of the magnitude of the localized spins, and they 
may be approximated by spins 1/2 \cite{Zang}. 
Noting that the double-exchange 
FM term Eq.\ (\ref{FMterm}) is insensitive to the 
sign of $J$,  
it may be readily verified that the derivation 
of the effective hamiltonian of Eq.\ (\ref{Heff})
carries over to this case with minor modifications. 
This determines a quantum order-disorder transition  
from a FM to a disordered PM phase, with a critical 
line for $|J_{c}|$ given by Eq.\ (\ref{fullJc}). 

To plot the critical line, we follow the analysis of 
Sec.\ III and use available numerically determined 
transition points for the $J < 0$ KLM. 
Yunoki {\it et al.} \cite{Yunoki} 
determine the FM-PM transition for classical spins 
via Monte Carlo, and for quantum spins 3/2 via the 
DMRG. The resulting transition lines, with coupling 
$J$ correspondingly scaled, are very close, and their 
points may be used within our spin 1/2 approximation. 
In Fig.\ 5 we plot the dimensionless parameter 
$Ja/v_{F}$ against $J$ for numerically determined 
points. The straight line of best fit gives very good 
agreement with the points, and as in Sec.\ III, 
determines the critical line
\beqa
-J_{c} = \frac{0.7 \sin(\pi n/2)}{1 - \sin(\pi n/2)}\, .
\label{Jcfm}
\eeqa
The resulting phase diagram is given in Fig.\ 6. 
The critical line diverges close to half-filling, 
and differs from the $J > 0$ KLM for which the line 
remains finite. 
We have not included the phase separated region 
identified by Yunoki {\it et al.} \cite{Yunoki} in 
Fig.\ 6. Phase separation is observed in the classical 
spin simulation in the PM region from $J_{c} = 4$. 
It is not observed in the quantum simulation until 
$J = 6$, and then occurs away from the FM-PM 
transition closer to half filling. Any phase 
separation involves strongly localized  
electrons, and on-site localization is not described 
well by our bosonization of the conduction 
band (cf. Sec.\ II).

\begin{figure}
\centering
\epsfig{file=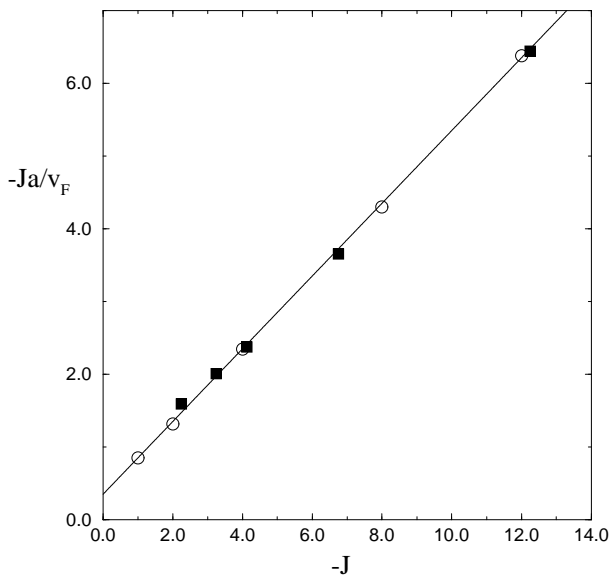,height=9cm}
\caption{Plot of the dimensionless parameter $-Ja/v_{F}$, 
which characterizes double-exchange ordering in 
the $J < 0$ KLM, against coupling $J$ for 
numerically determined FM-PM transition points: 
Open circles are results on classical 
localized spins using Monte Carlo on systems up 
to 40 sites from Ref.\ [8]; filled squares are 
DMRG results on a 16 site chain for quantum 
spins 3/2, and a correspondingly normalized 
coupling, from Ref.\ [8]. The straight line of 
best fit gives very good agreement with all points.}
\end{figure}

\begin{figure}[tb]
\centering
\epsfig{file=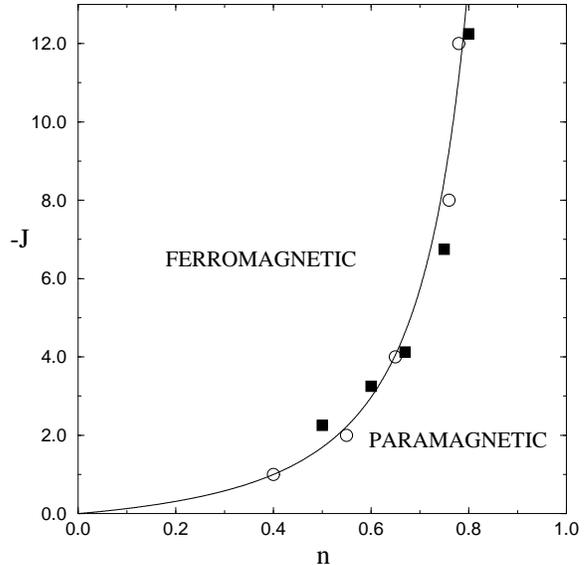,height=9cm}
\caption{Phase diagram for the 1D KLM with a  
ferromagnetic coupling $J < 0$. 
The critical line is from 
Eq.\ (\ref{Jcfm}), and uses the line of Fig.\ 5. 
Numerically determined transition points are as in 
Fig.\ 5. Properties of the localized 
spins close to criticality and at weak-coupling 
are as for the $J > 0$ KLM. The phase 
separated region identified in Ref.\ [8] for the 
classical spins from $J_{c} = 4$ into the 
paramagnetic phase is not shown. Phase separation 
is observed in Ref.\ [8] in the quantum simulation 
only at stronger couplings, and away from the FM-PM 
transition closer to half filling.}
\end{figure}

The effective range $\alpha$ of the double-exchange 
interaction on the transition line may be determined 
as in Sec.\ III. For an exponential cut-off function, 
we find $\alpha/a = \sqrt{0.7/J}$. As 
a function of filling, this relation may be used  
together with the critical line Eq.\ (\ref{Jcfm}) 
to plot $\alpha/a$ against $n$ as shown in Fig.\ 4.
The vanishing of the effective range close to half 
filling is the reason the critical line diverges. 

The different filling dependence of $\alpha$ for $J > 0$ and 
$J < 0$ is shown in Fig.\ 4. Different effective ranges 
$\alpha$ for the double-exchange interaction for different signs 
of the coupling is due to the different 
infinite $|J|$ symmetries of the sites containing 
localized conduction electrons. For a FM coupling,  
this is a triplet with energy  $-|J|/4$, while for 
Kondo couplings $J > 0$ the on-site symmetry is singlet with the 
lower energy $-3J/4$. The gain in 
energy for double-exchange per site and per conduction 
electron in either case is $-|J|/4$. 
For $J < 0$, the system thus gains just as 
much energy from double-exchange as it does by 
forming localized triplets, and the triplets have 
minimal effect for a non-vanishing hopping. 
The effective range $\alpha$ vanishes smoothly as 
the number of excess localized spins declines: $\alpha 
\rightarrow 0$ as $n \rightarrow 1$. For $J > 0$ the 
situation is far more complex. The on-site singlet 
energy is lower than the gain for double-exchange, 
and there is a complicated co-existence between the 
two effects. (This is why the exact solution 
for the KLM with one conduction electron is complex 
for $J > 0$, whereas it is trivial for $J < 0$ 
\cite{one-electron}.) Fig.\ 4 indicates a saturation 
$\alpha \approx a$ as $n \rightarrow 1$ for $J > 0$, 
in contrast to the $J < 0$ behavior. We leave as an  
outstanding problem the reason for this: An accurate 
determination requires a detailed description of localized 
Kondo singlet formation on a par with the double-exchange 
ordering and weak-coupling spin-flip scattering which 
have been the focus here.

\section{DISCUSSION AND CONCLUSIONS}

In this paper, we have described double-exchange FM ordering 
in the partially-filled 1D KLM, and 
the destuction of the FM phase by spin-flip disorder scattering.  
At weak-coupling deep in the disordered PM phase, the scattering  
determines RKKY-like correlations in the localized spins. 
Kondo singlet formation has been taken into account 
indirectly, via an effective range for the double-exchange 
interaction. The effective range was identified as follows: 
A length $\alpha \gtrsim {\cal O}(k_{F})^{-1}$ originates  
in bosonization as the minimum wavelength for density 
fluctuations which satisfy bosonic commutation relations.  
Bosonization describes fluctuations beyond $\alpha$, and 
keeps the conduction electrons finitely delocalized 
over lengths below $\alpha$. The electrons preserve their 
spin over this range. We showed in Sec.\ II that 
this finite delocalization may be identified with the  
length for coherent conduction electron hopping, and  
measures the effective range of the double-exchange 
FM ordering induced on the localized spins by the electrons
(cf.\ Eq.\ (\ref{FMterm}) and Fig.\ 1). The reason this 
works is that double-exchange is conceptually a simple 
interaction. It reflects only the tendency for hopping 
electrons to preserve their spin, 
as they move to screen the more numerous localized spins.
Double-exchange is characterized by the dimensionless 
factor $Ja/v_{F}$, with $v_{F}$ the conduction electron 
Fermi velocity and $a$ the lattice spacing. 

We obtained a FM double-exchange interaction term Eq.\ (\ref{FMterm}) 
between the localized spins. The term was derived
using a unitary transformation and is non-perturbative. 
This contrasts with other interactions derived for the localized spins 
in the KLM, such as the RKKY interaction, which are perturbative. 
The unitary transformation generates an effective 
hamiltonian Eq.\ (\ref{Heff}) for the localized spins. 
The competing affects on the spin ordering are made 
manifest in the effective hamiltonian. The competing effects 
are double-exchange ordering at stronger coupling, and 
spin-flip disorder processes involving nearly free  
electrons at weak-coupling. 
The transition from a double-exchange 
ordered FM phase to a quantum disordered PM phase 
was then shown in Sec.\ III to be the quantum order-disorder 
transition of the transverse-field Ising chain 
(cf.\ Eq.\ (\ref{Jc})). 
This describes double-exchange ordered regions 
of localized spins being destroyed as the electrons 
become weakly-bound, and become free to move and scatter 
along the chain. As the coupling $J$ is lowered, the 
transition is signalled by a continuously vanishing 
spontaneous magnetization Eq.\ (\ref{Mo}), and a breakdown 
in long-range correlations between the localized spins 
Eq.\ (\ref{Cx}). The phase diagram is given in Fig.\ 3. 
Well below the critical line, 
no remnants of the ordering remain, and 
the effective hamiltonian describes dominant correlations 
in the localized spins at $2k_{F}$ of the conduction band.

Spin disorder 
occurs through forward and backscattering spin-flip 
processes between the electrons and the localized 
spins. We identified interesting properties 
resulting from an incommensurate modulation of the 
backscattering momentum transfer with respect to the 
underlying lattice of localized spins: For 
incommensurate fillings, the conduction band has a 
competing periodicity with respect to the spin chain, 
and the electrons are unable to totally order, or 
totally disorder the spin chain at criticality. This 
leaves anomalous regions of double-exchange ordered 
localized moments close to criticality in the PM 
phase, as only a quasi-commensurate fraction of the 
electrons become weakly-bound at the transition. 
Similarly, there remain anomalous disordered regions close 
to criticality in the FM phase. The anomalous regions are 
very dilute, but they dominate the low-energy behavior of 
the localized spins. The magnetization Eq.\ (\ref{M}) 
is highly singular 
for a finite range of couplings about the critical line:  
The magnetization has a continuously variable 
power law exponent, and the susceptibility is infinite for 
a finite range of couplings even in the PM phase (cf.\ 
Eq.\ (\ref{chiT})). 

We considered in Sec.\ IV the effect of 
conduction electron interactions 
on the FM-PM transition, and found that double-exchange is 
characterized by the dimensionless factor 
$Ja\sqrt{v_{F}/v_{\sig}^{3}}$ for repulsive interactions, 
where $v_{\sig}$ is the conduction electron 
spin velocity. This factor 
is enhanced (cf.\ Eq.\ (\ref{52})) 
over the factor $Ja/v_{F}$ characterizing 
double-exchange with no interaction between the conduction 
electrons. This pushes the critical line to lower values 
of the coupling $J$, 
and for infinitely strong repulsive interactions 
FM occupies the entire phase 
diagram for $J \neq 0$. The reason for this behavior is 
that for infinite repulsive interactions, the double-exchange 
ordered regions are prevented from interfering, and the 
spin-flip disorder processes are ineffective. 

Since Kondo singlet formation is taken into account 
in our description only indirectly, our method 
extends also to the KLM with a FM $J < 0$ coupling, 
and a FM-PM transition of the same class as 
$J > 0$ was identified. The phase diagram for $J < 0$ is 
given in Fig.\ 6. The difference between $J > 0$ and   
$J < 0$ is in the effective range $\alpha$ of the 
double-exchange interaction, and is due to the different 
energies for on-site triplets when $J < 0$, to on-site Kondo 
singlets when $J > 0$ (cf.\ Sec.\ IV).
The effective range $\alpha$, which enters bosonization as a 
finite  
but unknown length, was determined at the FM-PM transition 
by using our critical line 
Eq.\ (\ref{fullJc}), together with numerically determined  
transition points 
\cite{Troyer,Tsunetsugu,Moukouri95,Caprara,Yunoki}. 
We found that $\alpha/a \propto 
1/\sqrt{|J|}$, in agreement with a simple characterization 
of double-exchange at small conduction band fillings 
(cf.\ Sec.\ II), and in agreement with an exact result 
\cite{one-electron} at vanishing filling. (Recall that the 
coupling $J$ is measured in units of hopping $t$). The 
proportionality constants are different for different 
signs of the coupling, and are fixed by a best fit to available 
numerically determined FM-PM transition points (Figs.\ 2 and 5). 
In Fig.\ 4 we plotted the 
corresponding filling dependence of $\alpha$ on the 
transition line:  
$\alpha \rightarrow 0$ as $n \rightarrow 1$ for $J < 0$, 
but remains finite for $J > 0$. This has a significant 
effect on the phase diagrams. For $J > 0$ (Fig.\ 3) the 
critical line remains finite as $n \rightarrow 1$, while 
for $J < 0$ (Fig.\ 6) the critical line diverges approaching 
half-filling. (Note that we do not consider the half-filled KLM. Indeed 
double-exchange FM is absent if the number of conduction electrons 
equals the number of localized spins.) 

The transition we identified is generic to partially-filled 
spin 1/2 KLMs, at least in 1D.
Our use of bosonization 
prevents us from anything more than speculation on 
the FM-PM transition in higher dimensional KLMs. 
We note only that (i) double-exchange is not restricted 
to 1D \cite{Zener,Anderson}, and should be considered 
in any discussion of partially-filled KLMs in higher 
dimensions. (ii) Numerical work \cite{Yunoki} on the KLM 
with a FM coupling does present a FM-PM 
transition in higher dimensions, which is very similar to 
the transition in the 1D case. 

We conclude with a simple 
physical picture, suggested to us by our results, 
which underlies the generic ground-state transition. 
At small fillings in the FM phase,  
spin 1/2 Kondo lattices form a 
gas of spin polarons, with each electron dressed 
by a cloud of ordered localized spins. The spatial extent 
of the polarization cloud is the effective range $\alpha$ for 
the double-exchange interaction. For $J > 0$ 
the localized spins tend to align opposite to the spin of the 
conduction electron. For $J < 0$ they tend 
to align parallel to the electron spin. 
As the coupling is lowered, the polarization clouds 
gradually extend and begin to interfere. The interference  
causes spin-flip disorder processes, which eventually destroy 
the FM order: The spin-flip  
processes free the electrons from their 
clouds of polarized localized spins, and this signals 
the onset of the FM-PM phase transition. At couplings just 
below the transition in the PM phase, 
the electrons are nearly free, and move through 
the system. They scatter from the localized spins as they move, 
and the spin chain is disordered. At weak-coupling, 
the localized spins retain dominant correlations at $2k_{F}$
of the conduction electrons, superimposed on an incoherent 
background. This reflects the momentum transferred from the 
conduction band to the spin chain in 
backscattering interactions, together with 
incoherent forward scattering.

\section*{Acknowledgments}
The authors thank P. W. Anderson, A. R. Bishop, D. S. Fisher, 
A. Muramatsu, S. A. Trugman, and J. Voit for useful discussions. 
This work was supported by the Australian Research Council.


\begin{references}
\bibitem{Lee}{See, for example, P. A. Lee, T. M. Rice,
 J. W. Serene, L. J. Sham, and J. W. Wilkins, Comments
 Cond. Mat. Phys. {\bf 12}, 99 (1986).}
\bibitem{Zang}{J. Zang, H. R\"{o}der, A. R. Bishop, and 
 S. A. Trugman, J. Phys.: Condens. Matter {\bf 9}, 
 L157 (1997).}
\bibitem{Honner}{G. Honner and M. Gul\'{a}csi, Phys. Rev. 
 Lett. {\bf 78}, 2180, (1997).}
\bibitem{Troyer}{M. Troyer and D. W\"{u}rtz, Phys. Rev. B
 {\bf 47}, 2886 (1993).}
\bibitem{Tsunetsugu}{H. Tsunetsugu, M. Sigrist, and
 K. Ueda, Phys. Rev. B {\bf 47}, 8345 (1993).}
\bibitem{Moukouri95}{S. Moukouri and L. G. Caron, Phys. Rev. B
 {\bf 52}, R15723 (1995).}
\bibitem{Caprara}{S. Caprara and A. Rosengren, Europhys. 
 Lett. {\bf 39}, 55 (1997).}
\bibitem{Yunoki}{S. Yunoki, J. Hu, A. L. Malvezzi, A. Moreo, 
 N. Furukawa, and E. Dagotto, cond-mat/9706014. See also 
 E. Dagotto {\it et al.} cond-mat/9709029.}
\bibitem{Doniach}{S. Doniach, Physica {\bf 91B}, 231, 
 (1977).}
\bibitem{Yanagisawa}{See, for example, T. Yanagisawa and 
 M. Shimoi, Int. J. Mod. Phys. B {\bf 10}, 3383 (1996).}
\bibitem{one-electron}{M. Sigrist, H. Tsunetsugu, and
 K. Ueda, Phys. Rev. Lett. {\bf 67}, 2211 (1991).}
\bibitem{strong-coupling}{M. Sigrist, H. Tsunetsugu,
 K. Ueda, and T. M. Rice, Phys. Rev. B {\bf 46}, 13838 
 (1992).}
\bibitem{Anderson}{P. W. Anderson and H. Hasegawa, 
 Phys. Rev. {\bf 100}, 675 (1955).}
\bibitem{Yanagisawa94}{T. Yanagisawa and K Harigaya, 
 Phys. Rev. B {\bf50}, 9577 (1994).}
\bibitem{Moukouri96}{S. Moukouri, L. Chen, and L. G. Caron,
 Phys. Rev. B {\bf 53}, R488 (1996).}
\bibitem{Voit}{See, for example, J. Voit, Rep. Prog. Phys. 
 {\bf 57}, 977 (1994).}
\bibitem{Tomonaga}{S. Tomonaga, Prog. Theor. Phys. {\bf 5},
 544 (1950).}
\bibitem{Schick}{M. Schick, Phys. Rev. {\bf 166}, 404 (1968).}
\bibitem{Hirsch}{J. E. Hirsch, Phys. Rev. B {\bf 30},
 5383 (1984).}
\bibitem{Emery}{V. J. Emery and S. Kivelson,
 Phys. Rev. B {\bf 46}, 10812 (1992).}
\bibitem{ZKE}{O. Zachar, S. A. Kivelson, and V. J. Emery, 
 Phys. Rev. Lett. {\bf 77}, 1342 (1996).}
\bibitem{Zener}{C. Zener, Phys. Rev. {\bf 82}, 403 (1951).}
\bibitem{Makhankov}{See, for example, V. G. Makhankov, 
 {\it Soliton Phenomenology}, Mathematics and Its Applications 
 (Soviet Series) Vol. 33 (Kluwer, Dordrecht, 1989).}
\bibitem{Holstein}{T. Holstein, Ann. Phys. (N.Y) {\bf 16}, 407 
 (1961).}
\bibitem{Jullien}{R. Jullien, J. N. Fields, and S. Doniach, 
 Phys. Rev. B {\bf 16}, 4889 (1977).}
\bibitem{Lieb}{E. Lieb, T. Schultz, and D. Mattis, Ann. Phys. 
 (N.Y.) {\bf 16}, 407 (1961).}
\bibitem{Pfeuty79}{P. Pfeuty, Phys. Lett {\bf 72A}, 245 (1979).}
\bibitem{Pfeuty70}{P. Pfeuty, Ann. Phys. (N.Y.) {\bf 57}, 
 79 (1970).}
\bibitem{Strong}{S. P. Strong and A. J. Millis, Phys. Rev. B 
 {\bf 50}, 9911 (1994).}
\bibitem{Satija89}{I. I. Satija and M. M. Doria, Phys. Rev. B 
 {\bf 39}, 9757 (1989).}
\bibitem{Satija90}{I. I. Satija, Phys. Rev. B 
 {\bf 41}, 7235 (1990).} 
\bibitem{Satija94}{I. I. Satija, Phys. Rev. B {\bf 49}, 
 3391 (1994).}
\bibitem{Fisher}{Daniel S. Fisher, Phys. Rev. Lett. {\bf 69},
 534 (1992); Phys. Rev. B {\bf 51}, 6411 (1995).}
\bibitem{Griffiths}{R. B. Griffiths, Phys. Rev. Lett.
 {\bf 23}, 17 (1969).}
\bibitem{caveat1}{For certain quasi-commensurate fillings, 
 for example $n = 1/2$, the spontaneous magnetization still 
 grows continuously, but the critical exponent may 
 be reduced to its value 1/8 as in the constant 
 transverse-field Ising chain. See Ref.\ [28].} 
\bibitem{caveat2}{Note that some features of the transition 
 in the random transverse-field Ising chain are, however,  
 loosely similar to those of first order transitions in 
 random classical systems. See Ref.\ [33].}
\bibitem{Wagner}{See, for example, M. Wagner, {\it Unitary 
 Transformations in Solid State Physics}, Modern Problems 
 in Condensed Matter Sciences Vol. 15 
 (Elsevier, Amsterdam, 1986).}
\bibitem{Solyom}{See, for example, J. S\'{o}lyom, Adv. 
 Phys. {\bf 28}, 201 (1979), and references therein.}
\bibitem{Schulz}{H. J. Schulz, Phys. Rev. Lett. {\bf 64}, 
 2831 (1990); Int. J. Mod. Phys. B {\bf 5}, 57 (1991).}
\end{references}
\end{document}